\documentclass[journal]{IEEEtran}[12pt]
\ifCLASSINFOpdf
\usepackage[pdftex]{graphicx}
\graphicspath{{../pdf/}{../jpeg/}}
\DeclareGraphicsExtensions{.pdf,.jpeg,.png}
\else
\usepackage[dvips]{graphicx}
\graphicspath{{../eps/}}
\DeclareGraphicsExtensions{.eps}
\fi\usepackage{graphicx}

\usepackage{graphics}
\usepackage{epsfig}
\usepackage{epstopdf}
\usepackage{stfloats}
\usepackage[cmex10]{amsmath}
\usepackage{algorithmic}
\usepackage[ruled]{algorithm2e}
\usepackage{indentfirst}
\usepackage{array}
\usepackage{mdwmath}
\usepackage{mdwtab}
\usepackage{graphicx}
\usepackage{subfigure}
\usepackage{color}
\usepackage{amsfonts,amssymb}
\usepackage{multirow}
\usepackage{diagbox}
\usepackage{caption}
\usepackage{verbatim}
\usepackage{float}
\usepackage{graphicx}
\usepackage{bm}
\usepackage{bbding}
\usepackage{tabu} 
\usepackage{multirow} 
\usepackage{multicol} 
\usepackage{multirow} 
\usepackage{float} 
\usepackage{makecell}
\usepackage{booktabs} 
\usepackage [mathscr] {euscript}
\usepackage{threeparttable}
\usepackage{hyperref} %

\usepackage{lineno} 

\usepackage{amsfonts,amsthm,array} 

\usepackage{amsfonts,amsthm,array} 
\makeatletter
\renewcommand{\maketag@@@}[1]{\hbox{\m@th\normalsize\normalfont#1}}
\makeatother

\SetKwRepeat{Do}{do}{while}

\begin{document}

\title{On Secure EKF-enhanced UAV-ISAC Systems \thanks{Manuscript received.}}

\author{Hongjiang~Lei, 
	Heng~Jin,
	Ki-Hong~Park, 
	Jia~Ye, 
	Liang~Yang, 
	Gaofeng~Pan, 
	and
	Yun~Li
\thanks{This work was supported by the National Key Research and Development Program of China under Grant 2024YFC3306801, National Natural Science Foundation of China under Grant 62571045, and Natural Science Foundation of Chongqing under Grant CSTB2025NSCQ-LZX0053. (Corresponding author:  \textit{Hongjiang~Lei}.)}
\thanks{Hongjiang~Lei and Heng~Jin are with School of Communications and Information Engineering, Chongqing University of Posts and Telecommunications, Chongqing 400065, China (e-mail: leihj@cqupt.edu.cn, cquptjh@163.com).}
\thanks{Ki-Hong~Park is with the CEMSE Division, King Abdullah University of Science and Technology (KAUST), Thuwal 23955-6900, Saudi Arabia (e-mail: kihong.park@kaust.edu.sa).}
\thanks{Jia~Ye is with State Key Laboratory of Power Transmission Equipment Technology, School of Electrical Engineering, Chongqing University, Chongqing 400044, China (e-mail: jia.ye@cqu.edu.cn).}
\thanks{Liang~Yang is with the College of Computer Science and Electronic Engineering, Hunan University, Changsha 410082, China (e-mail: liangy@hnu.edu.cn).}
\thanks{Gaofeng~Pan are with the School of Cyberspace Science and Technology, Beijing Institute of Technology, Beijing 100081, China (e-mail: gfpan@bit.edu.cn).}
\thanks{Yun~Li is with Chongqing Key Laboratory of Mobile Communications Technology, Chongqing 400065, China (e-mail: liyun@cqupt.edu.cn).}
}

\maketitle
\begin{abstract}
	Integrated sensing and communication (ISAC) has emerged as a promising key technology for future wireless networks, enabling the efficient coordination of sensing and communication functions within limited resources. This work investigates a secure ISAC system assisted by an uncrewed aerial vehicle (UAV). By incorporating the extended Kalman filter (EKF), the proposed system is capable of delivering communication services to legitimate users while simultaneously jamming eavesdroppers and performing joint prediction and tracking of the trajectories of both legitimate and illegitimate users. Considering practical constraints such as {sensing beamwidth}, transmit power, and UAV's propulsion energy consumption, the secrecy rate is maximized through the joint design of transmit beamforming and UAV trajectory. To tackle the resulting highly non-convex optimization problem, an efficient iterative algorithm is developed by integrating block coordinate descent, successive convex approximation, and EKF, thereby yielding a high-quality suboptimal solution. Extensive simulation results validate the superior performance of the proposed scheme compared to benchmarks.
\end{abstract}

\begin{IEEEkeywords}
Integrated sensing and communication,
uncrewed aerial vehicle, 
extended Kalman filter, 
physical-layer security.
\end{IEEEkeywords}

\section{Introduction}
\label{sec:Introduction}

\subsection{Background and Related Works}

\begin{table*}[!]
	\renewcommand{\arraystretch}{1.5}
	\caption{ Related Works on EKF-aided ISAC Systems.}
	\label{table1}
	\centering
	\begin{tabular}{c|c|c|c|c|c|c}  
		\Xhline{1.2pt}
		{\textbf{Reference}}&{\textbf{UAV}}&{\textbf{Antenna}}&{\textbf{PLS }}&{\textbf{\makecell{Mobility of  GUs}}} & {\textbf{\makecell{Mobility of targets}}} & {\textbf{\makecell{Number of sensing target}}}\\	
		\hline
		\cite{MengK2024TVT}   &  & ULA &  &\checkmark &\checkmark  &Multiple    \\	
		\hline
		\cite{WangZ2024JTSP}   &  & UPA &  &\checkmark&\checkmark  &Multiple    \\	
		\hline
		\cite{LiuF2020TWC}   &  &ULA &  &\checkmark &\checkmark  &Multiple     \\
		\hline
		\cite{WuJ2023IOT}   & \checkmark & Single &  &\makecell &\makecell  &Single    \\
		\hline
		\cite{LiuP2023TVT}   &  &UPA  &\checkmark  & &\checkmark  & Single    \\
		\hline
		\cite{WuJ2023TVT}   & \checkmark & Single &\checkmark &\checkmark &\checkmark  & Single    \\
		\hline
		\cite{ZhongJ2025CL}   & \checkmark & ULA & &\checkmark &\checkmark  & Multiple   \\
		\hline
		\cite{PangX2024TWC}   & \checkmark & ULA & &\checkmark &\checkmark  & Single   \\
		\hline
		\cite{ZhouS2024TCom}   & \checkmark & ULA & &\checkmark &\checkmark  & Multiple    \\
		\hline
		\cite{JiangY2024CL}   & \checkmark & ULA & &\checkmark &\checkmark  & Single    \\
		\hline
		\cite{MaoW2025JSAC}   & \checkmark & UPA & &\checkmark &\checkmark  & Multiple   \\
		\hline
		\cite{JinH2026IoT}   & \checkmark & ULA & \checkmark &\checkmark &\checkmark  & Single   \\
		\hline
		\cite{WeiZ2024TWC}   & \checkmark & UPA &\checkmark & &\checkmark  & Single    \\
		\hline
		Our Work     & \checkmark & UPA & \checkmark & \checkmark&\checkmark & Multiple  \\
		\hline
		\Xhline{1.2pt}
	\end{tabular}
\end{table*}

Integrated sensing and communication (ISAC) has emerged as a key enabling technology for sixth-generation (6G) networks, as it unifies sensing and communication functionalities to enable efficient sharing of spectrum and hardware resources while enhancing overall system performance \cite{ZhangD2026Surveys, JabeenN2025IOT}. 
Owing to their high mobility, flexible deployment, and line-of-sight (LoS) communication advantages, uncrewed aerial vehicles (UAVs) can effectively extend wireless coverage and mitigate signal blockage caused by obstacles \cite{WuQ2021JSAC}.
Consequently, the integration of UAV and ISAC emerged as a prominent research direction in the 6G domain. By fully exploiting their synergistic benefits, UAV-ISAC enabled enhanced sensing–communication cooperation, thereby facilitating applications such as high-precision tracking and three-dimensional network deployment \cite{WuZ2025IoTMag, MengK2024WC, FeiZ2023Mag}.

In UAV-ISAC systems, the integration of sensing and communication allows UAVs to detect eavesdroppers via sensing and use this real-time feedback to adjust their communication strategies. For example, UAVs can actively avoid eavesdroppers or precisely target them with artificial noise to degrade the eavesdropping channel, which creates a closed-loop, sensing-assisted security mechanism \cite{YangH2025JKSUC}.
In \cite{LiuY2024TVT}, a dual-UAV secure communication framework was proposed, consisting of a source UAV and a cooperative jamming UAV. The source UAV employed ISAC to sense the locations of multiple eavesdroppers and shared this information with the jamming UAV, which then transmitted targeted jamming signals to maximize the secrecy rate (SR). 
Similarly, in \cite{SonC2025WCL}, a UAV-enabled ISAC system was considered, in which the UAV provided communication services to multiple ground users (GUs) while simultaneously sensing and jamming terrestrial eavesdroppers. In this work, the transmit/receive beamforming and UAV trajectory were jointly optimized to maximize both the system sum rate and the SR.
To further enhance sensing accuracy and secure transmission, additional techniques were integrated into the UAV-ISAC framework. 
In \cite{LeiH2025IOT}, a UAV-assisted integrated sensing, communication, and computing (ISCC) system was investigated through the joint optimization of offloading ratios, scheduling policies, transmit beamforming, and UAV trajectories. Subject to UAVs' transmit power and energy constraints, the total energy consumption of GUs was minimized while guaranteeing communication security and sensing performance. 
In \cite{XiuY2025TVT, WangY2026IoT}, the location of potential eavesdroppers were obtained by sending the sensing signals and the average SR (ASR) was maximized by jointly optimizing the aerial vehicle trajectory along with transmit and receive beamforming vectors. 
In \cite{ZhangJ2024TWC}, secure transmission in an intelligent reflecting surface (IRS)-assisted UAV-ISAC network was studied under unknown eavesdropper channel state information. In this setup, IRSs were deployed to support both communication and target sensing, thereby enabling secure and efficient transmission.
Despite these advances, the terrestrial targets in the aforementioned works were generally assumed to be stationary and known, or arbitrarily located but static within a given area. Consequently, target mobility, a critical factor in practical scenarios, remained largely unknown.

{To address the challenge of target mobility, several studies have incorporated predictive mechanisms into ISAC system design. Existing methods for handling target mobility can be broadly classified into data-driven approaches and model-driven approaches. Data-driven methods utilize deep learning networks to directly learn from historical reflected-sensing signals and to predict optimal beamformers in an end-to-end manner. For instance, the authors in \cite{WangZ2024JTSP} proposed deep learning-based predictive beamforming to mitigate the adverse effects of parameter estimation errors inherent in conventional two-phase schemes, introducing one-sided and two-sided predictive beamforming frameworks to enhance the achievable sum rate. \textit{The main advantages of data-driven methods lie in the ability to handle complex nonlinear relationships, and their avoidance of error propagation inherent in explicit estimation. However, they suffer from poor interpretability, strong dependence on large training datasets, and limited generalization due to the limited coverage of the training environments. In contrast, model-driven approaches are more classical and widely adopted in the literature. Their core idea is to establish a mobility model of the target and explicitly estimate and predict the target’s state parameters (e.g., position, velocity, and angle) from radar echo signals using tools such as the extended Kalman filter (EKF).} In \cite{MengK2024TVT}, the integration of intelligent omni-surfaces (IOSs) with sensing-assisted communications was investigated, and a two-phase ISAC protocol was proposed, consisting of a joint sensing–communication phase followed by a pure communication phase. By jointly optimizing transmit beamforming, IOS phase shifts, and the joint phase duration, closed-form approximations of the achievable rate under vehicle location uncertainty were derived, thereby improving the achievable rate while reducing transmit power requirements. Notably, this work employs a Kalman-like prediction-measurement framework without implementing a full EKF.  
Other representative model-driven works include \cite{LiuF2020TWC}, \cite{WuJ2023IOT}, and \cite{LiuP2023TVT}. 
In \cite{LiuF2020TWC}, an EKF framework was employed to track and predict vehicular motion parameters, thereby reducing beam tracking overhead and achieving a favorable tradeoff between sensing and communication performance. 
Building on this, \cite{WuJ2023IOT} utilized EKF-based prediction to track the motion parameters of GUs based on distance measurements derived from base-station sensing echoes. The UAV trajectory and user association were jointly optimized to synergistically enhance both sensing and communication performance. In the context of security, \cite{LiuP2023TVT} applied EKF to predict the motion states of mobile aerial eavesdroppers, and jointly designed radar waveforms and receiver beamformers based on the tracking information to improve system secrecy capacity and fairness. \textit{The main strengths of model-driven approaches are their clear physical interpretability and well-understood behaviour. Nevertheless, their performance critically depends on the accuracy of both the motion and measurement models.}}

Furthermore, exploiting UAV mobility offers a promising solution to both enhance tracking performance and overcome the coverage limitations inherent in fixed terrestrial infrastructure. This inherent flexibility enables the system to adapt more effectively to diverse and dynamic operational scenarios.
A substantial body of research has integrated EKF-based prediction with UAV-ISAC systems to address the challenge of tracking mobile targets and users. In \cite{WuJ2023TVT}, a UAV-enabled EKF-assisted ISAC framework was investigated, wherein delay measurements extracted from ISAC echoes were exploited to track and predict the locations of unknown mobile legitimate users. A weighted nonconvex trajectory optimization problem was formulated to maximize the instantaneous SR against an eavesdropper moving along a predetermined trajectory. 
Similarly, in \cite{ZhongJ2025CL}, an EKF-based efficient tracking scheme was proposed to estimate mobile user locations, which were subsequently utilized for joint beamforming design and UAV trajectory optimization to maximize the average sum rate while enhancing sensing accuracy.
Beyond user tracking, EKF-based techniques have been extended to broader sensing and communication objectives. 
In \cite{PangX2024TWC}, vehicular state information was extracted from reflected signals and employed for vehicle state tracking and prediction using EKF. Based on real-time communication and sensing performance, the system selectively transmitted either ISAC beams or pure communication beams to satisfy asymmetric sensing–communication requirements and balance their performance tradeoff. 
In \cite{ZhouS2024TCom}, EKF was leveraged for accurate target state estimation and prediction, and a time-aided beamforming strategy was developed to improve communication throughput in both single-target and multi-target scenarios.
Optimizing UAV trajectories to support mobile target tracking has also received considerable attention. 
In \cite{JiangY2024CL}, UAV trajectory optimization was investigated for tracking mobile targets and communicating with onboard devices, where the weighted sum of the posterior Cramér–Rao bounds for relative position and velocity estimation errors was minimized. 
In \cite{MaoW2025JSAC}, EKF-based algorithms were employed to estimate mobile user locations, which were then combined with satellite-assisted positioning to predict channel distributions. The UAV trajectory, transmit beamforming vectors, and target transmission rates were jointly optimized to enhance overall system performance. 
A secure UAV-ISAC system was investigated in \cite{JinH2026IoT}, where EKF was used to estimate and predict the eavesdropper's state. Considering the maximum UAV velocity, transmit power, and propulsion energy constraints, the achievable uplink rate was maximized via joint transmit beamforming and UAV trajectory optimization. 
Moreover, the scope of EKF-assisted UAV systems has been extended to encompass integrated sensing, navigation, and communication. 
In \cite{WeiZ2024TWC}, a UAV-assisted EKF-enabled integrated sensing, navigation, and communication system was studied, in which online navigation was performed to minimize the distance to a predefined destination while accounting for channel-prediction errors and secure communication quality-of-service (QoS) constraints.
The representative works on EKF-aided ISAC systems are summarized in Table \ref{table1}.

\subsection{Motivation and Contributions}

Owing to the dominant LoS propagation characteristics of air-to-ground (A2G) channels, UAV-ISAC systems are capable of achieving significantly enhanced system performance. By further integrating ISAC with physical-layer security (PLS) techniques and EKF-based target tracking, the inherent advantages of UAVs, such as flexible deployment, high mobility, enhanced security, and low power consumption, can be more fully exploited. Within this framework, UAVs can communicate with GUs while simultaneously leveraging radar echoes to localize and predict the motion of mobile users and eavesdroppers. This enables the design of targeted communication and jamming strategies, thereby facilitating more secure and efficient transmissions under constrained resources.
Nevertheless, existing studies in this domain have either largely overlooked communication security considerations or focused exclusively on tracking and predicting a single target. To bridge these gaps, this work investigates a secure UAV-ISAC system enabling simultaneous tracking of both legitimate and illegitimate mobile targets. The main contributions of this work are summarized as follows.

\begin{enumerate}
	
	\item This work investigates a secure UAV-ISAC system in which an aerial base station transmits signals to both legitimate users and eavesdroppers. This enables simultaneous communication with users, real-time sensing, and jamming of eavesdroppers, while also facilitating the prediction and tracking of their motion states. Subject to practical constraints, including maximum flight speed, transmit power, energy consumption, and {sensing beamwidth} requirements, the achievable SR is maximized by jointly optimizing the sensing beamforming vectors and the UAV trajectory.

	\item Although existing studies, such as \cite{PangX2024TWC}–\cite{MaoW2025JSAC}, have explored UAV-ISAC systems that employ EKF for mobile target prediction, eavesdroppers are not explicitly considered in their system models. Consequently, these approaches are not directly applicable to the scenario under investigation in this work. In contrast, this work explicitly incorporates communication security into the system design by introducing a mobile eavesdropper with an uncertain location, thereby aligning the model more closely with practical deployment scenarios. Moreover, an EKF-based framework is adopted to estimate the real-time locations of multiple mobile targets, and beamforming strategies are designed to jointly enhance system security and sensing performance.
	
\end{enumerate}

\subsection{Organization}

The remainder of this paper is structured as follows. Section \ref{sec:SystemModel} introduces the system model, the EKF-based approach, and the problem formulation. The joint transmit beamforming and UAV trajectory optimization algorithm is developed in Section \ref{sec:proposed solution}. Numerical simulation results are presented in Section \ref{sec:Simulation}. Concluding remarks are given in Section \ref{sec:Conclusion}.
The notations and symbols adopted throughout this paper are summarized in Table \ref{table2}.
\begin{table}[t]
	
	\renewcommand{\arraystretch}{0.3}
	\caption{{List of Notations.}}
	\begin{center} 
		{
			{\begin{tabular}{c| c }
					\Xhline{1.2pt}
					{\textbf{Notation}}   	& {\textbf{Description}}								\\
					\hline
					${{\bf{q}}_B}({{\bf{q}}_U},{{\bf{q}}_E})$              & Horizontal coordinate of $B(U,E)$	\\
					\hline
					${z_B}$         				& Altitude of $B$				\\
					\hline
					${\rho _0}$             	& Channel gain at reference distance 1 m\\
					\hline
					$\delta_t$					& 	Time slot length			\\
					\hline
					$\kappa$			& Match filter gain		\\
					\hline
					$d$&  Antenna spacing\\
					\hline
					$\lambda $ & Wavelength\\
					\hline
					$\varepsilon$						& Radar cross-section \\
					\hline
					${\bf{a}}$     				& Steering vector of transmitting arrays				\\
					\hline
					$\mu,\tau$     				& {\makecell{Doppler shift/time delay of \\the sensing signal}}				\\
					\hline
					${\theta _{BU}},{\phi _{BU}}$				& {\makecell{Azimuth and elevation angles \\between $B$ and $U$}} \\
					\hline
					${\theta _{BE}},{\phi _{BE}}$				& {\makecell{Azimuth and elevation angles \\between $B$ and $E$}} \\
					\hline
					${\bf{s}},{\bf{m}}$				& Target state/measurement vector \\
					\hline
					${\bf{g}},{\bf{h}}$  		& State transition/measurement equations\\
					\hline
					${\bf{Q}}_s,{\bf{Q}}_m$			&  {\makecell{Covariance matrix of the state \\transmission/measurement noise}}\\
					\hline
					${\bf{M}}$&  Covariance matrix of ${\bf{s}}$\\
					\hline
					${\bf{K}}$ & Kalman gain\\
					\Xhline{1.2pt}
			\end{tabular}}
		}
	\end{center}
	\label{table2}
\end{table}	

\section{System Model and Problem Formulation}
\label{sec:SystemModel}

\begin{figure}[t]
	\centering
	\includegraphics[width = 0.31 \textwidth]{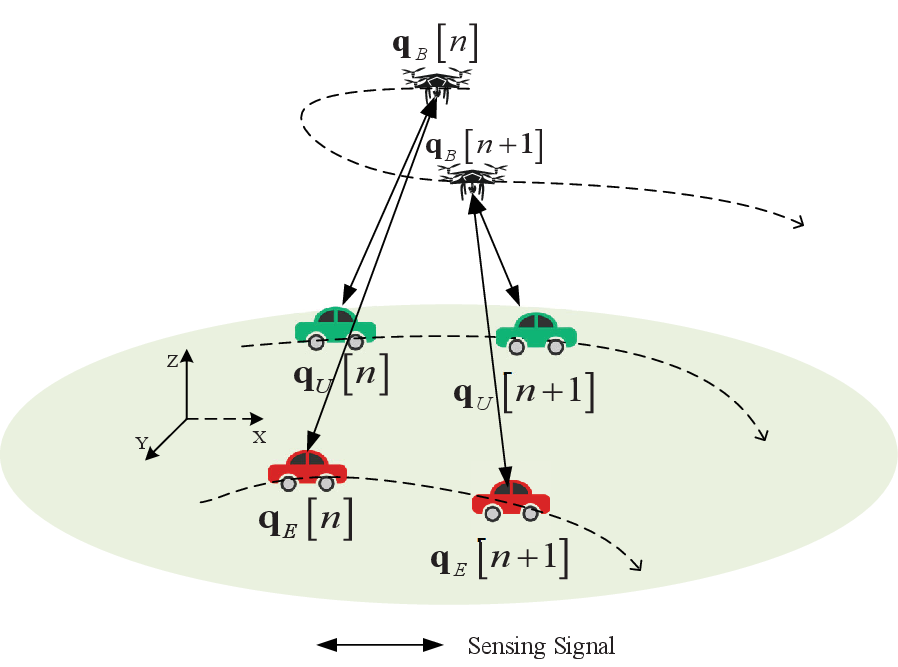}
	\caption{System model.}
	\label{figmodel}
\end{figure}

As illustrated in Fig.~\ref{figmodel}, this work investigates a classic PLS scenario, in which a full-duplex aerial base station ($B$) transmits confidential messages to a mobile GU ($U$), while an illegitimate GU ($E$) attempts to intercept the ongoing transmission. It is assumed that $B$ is equipped with two identical uniform planar arrays (UPAs) for transmitting radar signals and receiving signals with, each comprising $M = M_x \times M_y$ antenna elements, where $M_x$ and $M_y$ denote the numbers of elements along the $x$- and $y$-axes, respectively. Both $U$ and $E$ are assumed to be single-antenna nodes.

Like \cite{YaoY2025TCOM, YaoY2026TCOM}, to ensure analytical tractability, the total flight duration $T$ of $B$ is discretized into $N$ time slots, each with a duration of ${\delta _t} = T/N$. 
Both $U$ and $E$ move along predetermined ground trajectories, whose coordinates are denoted by ${{\bf{q}}_U}\left[ n \right] = {\left[ {{x_U}\left[ n \right],{y_U}\left[ n \right]} \right]^T}$ and ${{\bf{q}}_E}\left[ n \right] = {\left[ {{x_E}\left[ n \right],{y_E}\left[ n \right]} \right]^T}$, respectively, and their velocities by ${{\bf{v}}_U}\left[ n \right] = {\left[ {{v_{Ux}}\left[ n \right],{v_{Uy}}\left[ n \right]} \right]^T}$ and ${{\bf{v}}_E}\left[ n \right] = {\left[ {{v_{Ex}}\left[ n \right],{v_{Ey}}\left[ n \right]} \right]^T}$, respectively.

To enable efficient dual-functional operation, $B$ transmits communication signals to $U$ for data delivery while concurrently leveraging the same signals to predict and track the position of $U$. In parallel, $B$ actively transmits dedicated sensing signals to estimate the location of $E$, thereby facilitating the generation of intentional interference to degrade $E$'s reception.

\subsection{Communication and Sensing Model}

In the $n$-th time slot, the transmitted signal of $B$ is expressed as
\begin{align}
	{\bf{x}}\left( {n,t} \right) = {{\bf{w}}_U}\left[ n \right]{s_U}\left( {n,t} \right) + {{\bf{w}}_E}\left[ n \right]{s_E}\left( {n,t} \right), 
\end{align}
where 
${s_U}\left( {n,t} \right)$ and ${s_E}\left( {n,t} \right)$ 
denote the unit-power information-bearing signal for $U$ and the unit-power sensing signal for $E$, respectively, 
${{\bf{w}}_U}\left[ n \right]$ and ${{\bf{w}}_E}\left[ n \right]$ are the precoding vectors for $s_U$ and $s_E$, respectively.

{
Following \cite{WeiZ2024TWC, LyuZ2023TWC}, the LoS model is a reasonable approximation for the links between $B$ and ground targets in rural areas with little blockage or when $B$ flies at a sufficiently high altitude. Hence, the A2G links from $B$ to a terrestrial receiver $\chi$ ($\chi \in \{U, E\}$) are assumed to be LoS and are modelled as}
\begin{align}
	\mathbf{h}_{B\chi}\left[ n \right] = \sqrt{\rho_0 \, d_{B\chi}^{-2}\left[ n \right]} \, \mathbf{a}_{B\chi, T} \left( \theta_{B\chi}\left[ n \right], \phi_{B\chi}\left[ n \right] \right),
	\label{eq:channel_plural}
\end{align}
where 
$\rho_0$ denotes the path loss at a unit reference distance, 
$d_{B\chi}\left[ n \right]$ represents the distance between $B$ and $\chi$ in the $n$-th time slot, 
$\mathbf{a}_{B\chi, T} \left( \theta_{B\chi}\left[ n \right], \phi_{B\chi}\left[ n \right] \right)$ denotes the transmit steering vector of the antenna array, which is given by (\ref{antenna}) at the top of the next page, 
$d$ is the antenna spacing, 
$\lambda$ is the wavelength, 
and 
$\theta_{B\chi}\left[ n \right] \in [-\pi/2, \pi/2]$ and $\phi_{B\chi}\left[ n \right] \in [-\pi/2, 0]$ represent the azimuth and elevation angles of departure (AODs) from $B$ to $\chi$, respectively \cite{WeiZ2024TWC}.

\begin{figure*}[ht]
	\begin{align}
			{{\bf{a}}_{B\chi,T}}\left( {{\theta _{B\chi}}\left[ n \right],{\phi _{B\chi}}\left[ n \right]} \right) &= {\left( {1,{e^{ - \frac{{j2\pi d\cos \left( {{\phi _{B\chi}}\left[ n \right]} \right)\sin \left( {{\theta _{B\chi}}\left[ n \right]} \right)}}{\lambda }}}, \cdots ,{e^{ - \frac{{j2({M_x} - 1)\pi d\cos \left( {{\phi _{B\chi}}\left[ n \right]} \right)\sin \left( {{\theta _{B\chi}}\left[ n \right]} \right)}}{\lambda }}}} \right)^T} \nonumber\\
			&\otimes {\left( {1,{e^{ - \frac{{j2\pi d\cos \left( {{\phi _{B\chi}}\left[ n \right]} \right)\cos \left( {{\theta _{B\chi}}\left[ n \right]} \right)}}{\lambda }}}, \cdots ,{e^{ - \frac{{j2({M_y} - 1)\pi d\cos \left( {{\phi _{B\chi}}\left[ n \right]} \right)\cos \left( {{\theta _{B\chi}}\left[ n \right]} \right)}}{\lambda }}}} \right)^T}
		\label{antenna}
	\end{align}
	\hrulefill
\end{figure*}

Following \cite{WeiZ2024TWC}, it is assumed that $\chi$ achieves perfect time and frequency synchronization. Under this assumption, the received signal at $\chi$ in the $n$-th time slot is expressed as
\begin{align}
	r_\chi\left( t \right) &= \mathbf{h}_{B\chi}^H\left[ n \right] \mathbf{w}_U\left[ n \right] s_U\left( n, t \right) \nonumber\\
	&+ \mathbf{h}_{B\chi}^H\left[ n \right] \mathbf{w}_E\left[ n \right] s_E\left( n, t \right) + n_\chi\left( t \right),
	\label{eq:received_signal}
\end{align}
where $n_\chi\left( t \right) \sim \mathcal{CN}(0, \sigma_\chi^2)$ denotes the additive white Gaussian noise (AWGN) at $\chi$.
The signal-to-interference-plus-noise ratio (SINR) at $\chi$ is given by
\begin{align}
	\gamma_\chi\left[ n \right] = \frac{ \left| \mathbf{h}_{B\chi}^H\left[ n \right] \mathbf{w}_U\left[ n \right] \right|^2 }{ \psi_\chi \left| \mathbf{h}_{B\chi}^H\left[ n \right] \mathbf{w}_E\left[ n \right] \right|^2 + \sigma_\chi^2 },
	\label{eq:sinr}
\end{align}
where $0 \leq \psi_\chi \leq 1$ denotes the interference residual level\footnote{In this work, we consider a more practical scenario wherein, depending on the value of $\psi_\chi$, the interference can be either perfectly eliminated, partially suppressed, or remain non-eliminable. Specifically, $\psi_\chi = 0$ indicates perfect interference cancellation, $\psi_\chi = 1$ signifies no cancellation capability and $0 < \psi_\chi < 1$ corresponds to partial interference cancellation. {In this work, it is set as $\psi_E = 1$}.}.
Accordingly, the achievable rate at $\chi$ is expressed as
\begin{align}
	R_\chi\left[ n \right] = \log_2 \left( 1 + \gamma_\chi\left[ n \right] \right).
	\label{eq:rate}
\end{align}
The achievable SR of the considered system is then given by \cite{LeiH2025IOT, LiJ2026JSAC}
\begin{align}
	R_{\sec}\left[ n \right] = \left[ R_U\left[ n \right] - R_E\left[ n \right] \right]^+,
	\label{eq:secrecy_rate}
\end{align}
where $[x]^+ = \max\{x, 0\}$.

The horizontal flight energy consumption of $B$ is given by \cite{ZengY2019TWC}-\cite{YaoY2026TWC}:
\begin{align}
	{P_{\rm{hor}}}\left[ n \right] &= {P_0}\left( {1 + \frac{{3{{\left\| {{{\bf{v}}_{B}}\left[ n \right]} \right\|}^2}}}{{U_{{\rm{tip}}}^2}}} \right) + \frac{1}{2}{d_0}\rho sA{\left\| {{{\bf{v}}_{B}}\left[ n \right]} \right\|^3}\notag\\
	&+ {P_i}{\left( {\sqrt {1 + \frac{{{{\left\| {{{\bf{v}}_{B}}\left[ n \right]} \right\|}^4}}}{{4v_0^4}}}  - \frac{{{{\left\| {{{\bf{v}}_{B}}\left[ n \right]} \right\|}^2}}}{{2v_0^2}}} \right)^{1/2}}, \label{}
\end{align}
where 
${{{\bf{v}}_{B}}\left[ n \right]}  = \frac{{ {{{\bf{q}}_B}\left[ n \right] - {{\bf{q}}_B}\left[ {n - 1} \right]} }}{\delta_t }, (n > 1)$ denotes the average horizontal flight velocity of $B$, 
$P_0$ and $P_i$ represent the blade profile power and induced power under hovering conditions, respectively,
${U_{{\rm{tip}}}}$ denotes the rotor blade tip speed,
$v_0$ is the mean rotor-induced velocity in hover,
while $d_0$, $\rho$, $s$, and $A$ denote the fuselage drag ratio, air density, rotor solidity, and rotor disc area, respectively.

\subsection{Sensing Model}

According to \cite{ZhouS2024TCom,YuanW2021STSP}, the asymptotic orthogonality of steering vectors associated with different directions becomes increasingly pronounced as the number of antennas grows, thereby significantly mitigating mutual interference among reflected echoes. Nonetheless, in scenarios where targets are in close proximity and the asymptotic orthogonality condition alone proves inadequate for fully suppressing interference, digital beamforming techniques can be further utilized to reduce echo coupling. Building on this reasoning, the directionally phase-aligned echo received at $B$ is expressed as
\begin{align}
		{{\bf{r}}_{B \chi}}\left( {n,t} \right) &= {\beta _\chi}\left[ n \right]{e^{j2\pi {\mu _\chi}\left[ n \right]t}}{{\bf{a}}_{B\chi,R}}\left( {{\theta _{B\chi}}\left[ n \right],{\phi _{B\chi}}\left[ n \right]} \right)\notag\\
		&\times {\bf{a}}_{B\chi,T}^H\left( {{\theta _{B\chi}}\left[ n \right],{\phi _{B\chi}}\left[ n \right]} \right){{\bf{w}}_\chi}\left[ n \right]{s_\chi}\left( {n,t - {\tau _\chi}\left[ n \right]} \right) \notag\\
		&+ {{\bf{n}}_B}\left( {n,t} \right), \label{echo}
\end{align}
where 
${\beta _\chi}\left[ n \right] = \sqrt {\frac{{{\rho _0}}}{{d_{B \chi}^2\left[ n \right]}} \times \frac{\varepsilon }{{d_{B \chi}^2\left[ n \right]}}}$ denotes the reflection coefficient of the target \cite{LiuP2023TVT}, 
and $\varepsilon$ represents the radar cross section, 
which is assumed to be constant in this work, 
${\mu _\chi}\left[ n \right]$ and ${\tau _\chi}\left[ n \right]$ denote the Doppler frequency shift and time delay of the sensing signal, respectively, 
and ${\textbf{n}_B}\left( {n,t} \right)\sim\mathcal{CN}\left( {\textbf{0},\sigma _B^2{{\bf{I}}_{{M}}}} \right)$ represents the AWGN at $B$. 

It is worth noting that the clutter reflected from other scatterers in the environment is neglected here, since such clutter typically exhibits distinct reflection angles and Doppler frequencies compared with the target echoes, allowing existing clutter suppression techniques to effectively eliminate them \cite{WeiZ2024TWC}.
The SNR of the echo from $\chi$ at $B$ is expressed as
\begin{align}
	{\gamma _{\chi B}}\left[ n \right] = \frac{{\kappa {{\left| {{\beta _{\chi}}\left[ n \right]} \right|}^2}{{\left| {{\bf{a}}_{B{\chi}T}^H\left( {{\theta _{B\chi}}\left[ n \right],{\phi _{B\chi}}\left[ n \right]} \right){{\bf{w}}_{\chi}}\left[ n \right]} \right|}^2}}}{{\sigma _B^2}}, \label{echoSNR}
\end{align}
where $\kappa $ denotes the matched filter gain.

As described in \cite{WeiZ2024TWC}, by utilizing a matched filter, the parameters in (\ref{echo}), namely ${\mu _{\chi}}\left[ n \right]$, ${\tau _{\chi}}\left[ n \right]$, ${\theta _{B\chi}}\left[ n \right]$, and ${\phi _{B\chi}}\left[ n \right]$, can be determined through peak detection, and the expression (\ref{peak}) shown at the top of this page.
\begin{figure*}[ht]
	\begin{align}
		\left\{ {{{\tilde \tau }_{\chi}}\left[ n \right],{{\tilde \mu }_{\chi}}\left[ n \right],{{\tilde \theta }_{B\chi}}\left[ n \right],{{\tilde \phi }_{B\chi}}\left[ n \right]} \right\} = \arg \mathop {\max }\limits_{\mu ,\tau ,\theta ,\phi } \left| {\frac{1}{{{\delta _t}}}\int_0^{{\delta _t}} {{\bf{a}}_{B{\chi}R}^H\left( {{\theta _{B\chi}}\left[ n \right],{\phi _{B\chi}}\left[ n \right]} \right){{\bf{r}}_{B\chi}}} \left( t \right){s_\chi}^*\left( {t - {\tau _{\chi}}\left[ n \right]} \right){e^{ - j2\pi {\mu _{\chi}}\left[ n \right]t}}dt} \right|
		\label{peak}
	\end{align}
	\hrulefill
\end{figure*}

\subsection{Measurement Model}

The measurement model characterizes the relationship between observable target parameters and hidden state variables, serving as a fundamental tool for inferring target states. In particular, based on the relative positions between $B$ and $\chi$, the corresponding measurement model is formulated as \cite{WeiZ2024TWC}
\begin{subequations}
	\begin{align}
		&{\tau _{\chi}}\left[ n \right] = \frac{{2\sqrt {{{\left\| {{{\bf{q}}_{\chi}}\left[ n \right] - {{\bf{q}}_B}\left[ n \right]} \right\|}^2} + z_B^2\left[ n \right]} }}{c} + {z_{{\tau _{\chi}}}}\left[ n \right],\\
		&{\mu _{\chi}}\left[ n \right] = \frac{{2{{\left( {{{\bf{v}}_{\chi}}\left[ n \right] - {{\bf{v}}_B}\left[ n \right]} \right)}^T}\left( {{{\bf{q}}_{\chi}}\left[ n \right] - {{\bf{q}}_B}\left[ n \right]} \right){f_c}}}{{c\sqrt {{{\left\| {{{\bf{q}}_{\chi}}\left[ n \right] - {{\bf{q}}_B}\left[ n \right]} \right\|}^2} + z_B^2\left[ n \right]} }} + {z_{{\mu _{\chi}}}}\left[ n \right],\\
		&\sin {\theta _{\chi}}\left[ n \right] = \frac{{{x_{\chi}}\left[ n \right] - {x_B}\left[ n \right]}}{{\left\| {{{\bf{q}}_{\chi}}\left[ n \right] - {{\bf{q}}_B}\left[ n \right]} \right\|}} + {z_{\sin {\theta _{\chi}}}}\left[ n \right],\\
		&\sin {\phi _{\chi}}\left[ n \right] = \frac{{ - {z_B}\left[ n \right]}}{{\sqrt {{{\left\| {{{\bf{q}}_{\chi}}\left[ n \right] - {{\bf{q}}_B}\left[ n \right]} \right\|}^2} + z_B^2\left[ n \right]} }} + {z_{\sin {\phi _{\chi}}}}\left[ n \right], \label{sin}
	\end{align}
\end{subequations}
where ${f_c}$ denotes the carrier frequency, 
$c$ denotes the speed of light,
${z_{{\tau _{\chi}}}}\left[ n \right]$, ${z_{{\mu _{\chi}}}}\left[ n \right]$, ${z_{\sin {\theta _{\chi}}}}\left[ n \right]$, and ${z_{\sin {\phi _{\chi}}}}\left[ n \right]$ represent Gaussian measurement noises with zero mean and variances $\sigma _{{\tau _{\chi}}}^2\left[ n \right]$, $\sigma _{{\mu _{\chi}}}^2\left[ n \right]$, $\sigma _{\sin {\theta _{\chi}}}^2\left[ n \right]$, and $\sigma _{\sin {\phi _{\chi}}}^2\left[ n \right]$, respectively.  
It should be noted that 
$\sigma _{{\tau _{\chi}}}^2\left[ n \right]$, $\sigma _{{\mu _{\chi}}}^2\left[ n \right]$, $\sigma _{\sin {\theta _{\chi}}}^2\left[ n \right]$, and $\sigma _{\sin {\phi _{\chi}}}^2\left[ n \right]$
are inversely proportional to the echo SNR given in (\ref{echoSNR}), 
and 
are expressed as $\sigma _{{\tau _{\chi}}}^2\left[ n \right] = {{{a_1}} \mathord{\left/
		{\vphantom {{{a_1}} {{\gamma _{IB}}\left[ n \right]}}} \right.
		\kern-\nulldelimiterspace} {{\gamma _{\chi B}}\left[ n \right]}}$, $\sigma _{{\mu _{\chi}}}^2\left[ n \right] = {{{a_2}} \mathord{\left/
		{\vphantom {{{a_2}} {{\gamma _{IB}}\left[ n \right]}}} \right.
		\kern-\nulldelimiterspace} {{\gamma _{\chi B}}\left[ n \right]}}$, $\sigma _{{\theta _{\chi}}}^2\left[ n \right] = {{{a_3}} \mathord{\left/
		{\vphantom {{{a_3}} {{\gamma _{IB}}\left[ n \right]}}} \right.
		\kern-\nulldelimiterspace} {{\gamma _{\chi B}}\left[ n \right]}}$, and $\sigma _{{\phi _{\chi}}}^2\left[ n \right]{{ = {a_4}} \mathord{\left/
		{\vphantom {{ = {a_4}} {{\gamma _{IB}}\left[ n \right]}}} \right.
		\kern-\nulldelimiterspace} {{\gamma _{\chi B}}\left[ n \right]}}$, where the constant ${a_i},i = 1,2,3,4$ is defined as in \cite{LiuF2020TWC, LiuP2023TVT}, and depends on the system configuration as well as the adopted signal processing algorithms. Moreover, when $\sigma _{\sin {\theta _{\chi}}}^2\left[ n \right]$ and $\sigma _{\sin {\phi _{\chi}}}^2\left[ n \right]$ are sufficiently small, the angular estimates can be approximated using trigonometric identities as ${\sigma _{\sin {\theta _{\chi}}}}\left[ n \right] \approx \cos {\theta _{\chi}}\left[ n \right]{\sigma _{{\theta _{\chi}}}}\left[ n \right]$ and ${\sigma _{\sin {\phi _{\chi}}}}\left[ n \right] \approx \cos {\phi _{\chi}}\left[ n \right]{\sigma _{{\phi _{\chi}}}}\left[ n \right]$.

\subsection{State Evolution Model}

To achieve accurate tracking, a suitable state evolution model for the target is required in addition to the measurement model. Under the constant-velocity motion assumption \cite{WuJ2023TVT, MaoW2025JSAC}, the state evolution of $\chi$ is expressed as
\begin{subequations}
	\begin{align}
		{x_{\chi}}\left[ {n + 1} \right] &= {x_{\chi}}\left[ n \right] + {v_{\chi x}}\left[ n \right]{\delta _t} + {\eta _x}\left[ {n + 1} \right],\\
		{y_{\chi}}\left[ {n + 1} \right] &= {y_{\chi}}\left[ n \right] + {v_{\chi y}}\left[ n \right]{\delta _t} + {\eta _y}\left[ {n + 1} \right],\\
		{v_{\chi x}}\left[ {n + 1} \right] &= {v_{\chi x}}\left[ n \right] + {\eta _{vx}}\left[ {n + 1} \right],\\
		{v_{\chi y}}\left[ {n + 1} \right] &= {v_{\chi y}}\left[ n \right] + {\eta _{vy}}\left[ {n + 1} \right],
	\end{align}
\end{subequations}
where ${\eta _x}\left[ {n + 1} \right]$, ${\eta _y}\left[ {n + 1} \right]$, ${\eta _{vx}}\left[ {n + 1} \right]$, and ${\eta _{vy}}\left[ {n + 1} \right]$ denote the corresponding process noises, which are assumed to follow zero-mean Gaussian distributions with variances $\sigma _x^2$, $\sigma _y^2$, $\sigma _{vx}^2$, and $\sigma _{vy}^2$, respectively.

To perform local linearization of the nonlinear model, the formulated model is compactly rewritten as
\begin{align}
	\begin{array}{*{20}{l}}
		{{{\bf{s}}_{\chi}}\left[ n \right] = {\bf{g}}\left( {{{\bf{s}}_{\chi}}\left[ {n - 1} \right]} \right) + {\bm{\eta }}\left[ n \right]}\\
		{{{\bf{m}}_{\chi}}\left[ n \right] = {\bf{h}}_{\chi}\left( {{{\bf{s}}_{\chi}}\left[ n \right]} \right) + {{\bf{z}}_{\chi}}\left[ n \right],}
	\end{array}
\end{align}
where ${{\bf{s}}_{\chi}}\left[ n \right] = {\left[ {{x_{\chi}}\left[ n \right],{y_{\chi}}\left[ n \right],{v_{\chi x}}\left[ n \right],{v_{\chi y}}\left[ n \right]} \right]^T}$ and ${{\bf{m}}_{\chi}}\left[ n \right] = {\left[ {{\tau _{\chi}}\left[ n \right],{\mu _{\chi}}\left[ n \right],\sin {\theta _{\chi}}\left[ n \right],\sin {\phi _{\chi}}\left[ n \right]} \right]^T}$ denote the target state vector and the measurement vector, respectively, ${\bf{g}}\left( . \right)$ and ${\bf{h}}_{\chi}\left( . \right)$ are defined by the state transition equation and the measurement equation, respectively. The noise vectors ${\bm{\eta }}\left[ n \right] = {\left[ {{\eta _x}\left[ n \right],{\eta _y}\left[ n \right],{\eta _{vx}}\left[ n \right],{\eta _{vy}}\left[ n \right]} \right]^T}$ and ${{\bf{z}}_{\chi}}\left[ n \right] = {\left[ {{z_{{\tau _{\chi}}}}\left[ n \right],{z_{{\mu _{\chi}}}}\left[ n \right],{z_{\sin {\theta _{\chi}}}}\left[ n \right],{z_{\sin {\phi _{\chi}}}}\left[ n \right]} \right]^T}$ are independent of ${\bf{g}}\left( . \right)$ and ${\bf{h}}_{\chi}\left( . \right)$, and follow zero-mean Gaussian distributions with covariance matrices ${{\bf{Q}}_s} = {\rm{diag}}\left( {\sigma _x^2,\sigma _y^2,\sigma _{vx}^2,\sigma _{vy}^2} \right)$ and ${{\bf{Q}}_m} = {\rm{diag}}\left( {\sigma _{{\tau _{\chi}}}^2\left[ n \right],\sigma _{{\mu _{\chi}}}^2\left[ n \right],\sigma _{\sin {\theta _{\chi}}}^2\left[ n \right],\sigma _{\sin {\phi _{\chi}}}^2\left[ n \right]} \right)$, respectively.
The Jacobian matrices of ${\bf{g}}\left( . \right)$ and  ${\bf{h}}_{\chi}\left( . \right)$ are given by \cite{LiuP2023TVT, WuJ2023TVT, MaoW2025JSAC, WeiZ2024TWC}
\begin{align}
	{\bf{G}} = \frac{{\partial {\bf{g}}}}{{\partial {{\bf{s}}_{\chi}}}} = \left[ {\begin{array}{*{20}{c}}
			1&0&{{\delta _t}}&0\\
			0&1&0&{{\delta _t}}\\
			0&0&1&0\\
			0&0&0&1
	\end{array}} \right],
\end{align}
and 
 (\ref{jacobian of h}), shown at the top of this page,
{where 
$\Xi_1\left[ n \right] = 2{f_c}\left( \left( {{v_{\chi x}}\left[ n \right] - {v_{Bx}}\left[ n \right]} \right)d_{B\chi}^2\left[ n \right] - \Xi\left[ n \right] \left( {{x_{\chi}}\left[ n \right] - {x_B}\left[ n \right]} \right) \right)$,  
$\Xi_2\left[ n \right] = 2{f_c}\left( \left( {{v_{\chi y}}\left[ n \right] - {v_{By}}\left[ n \right]} \right)d_{B\chi}^2\left[ n \right] - \Xi\left[ n \right] \left( {{y_{\chi}}\left[ n \right] - {y_B}\left[ n \right]} \right) \right)$, 
and
$\Xi\left[ n \right] = \left( {{v_{\chi x}\left[ n \right]} - {v_{Bx}\left[ n \right]}} \right)\left( {{x_{\chi}\left[ n \right]} - {x_B\left[ n \right]}} \right) + \left( {{v_{\chi y}\left[ n \right]} - {v_{By}\left[ n \right]}} \right)\left( {{y_{\chi}\left[ n \right]} - {y_B\left[ n \right]}} \right)$.
}
\begin{figure*}[ht]
	\begin{align}
		{{\bf{H}}_{\chi}\left[ n \right]} &= {\left. {\frac{{\partial {{\bf{h}}_{\chi}}}}{{\partial {{\bf{s}}_{\chi}}}}} \right|_{{{\bf{s}}_{\chi}} = {{{\bf{\tilde s}}}_{\chi}}\left[ {n\left| {n - 1} \right.} \right]}} \nonumber \\
		&= \left. \left[ {\begin{array}{*{20}{c}}
				{\frac{{2\left( {{x_{\chi}}\left[ n \right] - {x_B}\left[ n \right]} \right)}}{{c{d_{B\chi}}\left[ n \right]}}}&{\frac{{2\left( {{y_{\chi}}\left[ n \right] - {y_B}\left[ n \right]} \right)}}{{c{d_{B\chi}}\left[ n \right]}}}&0&0\\
				{\frac{\Xi_1\left[ n \right]}{{cd_{B\chi}^3\left[ n \right]}}}&{\frac{\Xi_2\left[ n \right]}{{cd_{B\chi}^3\left[ n \right]}}}&{\frac{{2{f_c}\left( {{x_{\chi}}\left[ n \right] - {x_B}\left[ n \right]} \right)}}{{c{d_{B\chi}}\left[ n \right]}}}&{\frac{{2{f_c}\left( {{y_{\chi}}\left[ n \right] - {y_B}\left[ n \right]} \right)}}{{c{d_{B\chi}}\left[ n \right]}}}\\
				{\frac{{{{\left( {{y_{\chi}}\left[ n \right] - {y_B}\left[ n \right]} \right)}^2}}}{{{{\left\| {{{\bf{q}}_{\chi}}\left[ n \right] - {{\bf{q}}_B}\left[ n \right]} \right\|}^3}}}}&{\frac{{ - \left( {{x_{\chi}}\left[ n \right] - {x_B}\left[ n \right]} \right)\left( {{y_{\chi}}\left[ n \right] - {y_B}\left[ n \right]} \right)}}{{{{\left\| {{{\bf{q}}_{\chi}}\left[ n \right] - {{\bf{q}}_B}\left[ n \right]} \right\|}^3}}}}&0&0\\
				{\frac{{{z_B}\left[ n \right]\left( {{x_{\chi}}\left[ n \right] - {x_B}\left[ n \right]} \right)}}{{d_{B\chi}^3\left[ n \right]}}}&{\frac{{{z_B}\left[ n \right]\left( {{y_{\chi}}\left[ n \right] - {y_B}\left[ n \right]} \right)}}{{d_{B\chi}^3\left[ n \right]}}}&0&0
		\end{array}} \right] \right|_{{{\bf{s}}_{\chi}} = {{\bf{\tilde s}}_{\chi}}\left[ n \mid n-1 \right]}
		\label{jacobian of h}
	\end{align}
	\hrulefill
\end{figure*}

The procedure for state prediction and tracking are summarized as follows:
\begin{enumerate}
	\item Based on the state transition matrix, the priori predicted state is obtained as 
	\begin{align}
		{{\bf{\tilde s}}_{\chi}}\left[ {n\left| {n - 1} \right.} \right] = {\bf{g}}\left( {{{{\bf{\tilde s}}}_{\chi}}\left[ {n - 1} \right]} \right). \label{predicate_s}
	\end{align}
	\item By linearizing the state and observation equations, the state transition and measurement Jacobian matrices are obtained as
	\begin{align}
		{\bf{G}}\left[ {n - 1} \right] = {\left. {\frac{{\partial {\bf{g}}}}{{\partial {{\bf{s}}_{\chi}}}}} \right|_{{{\bf{s}}_{\chi}} = {{{\bf{\tilde s}}}_{\chi}}\left[ {n\left| {n - 1} \right.} \right]}}
	\end{align}
	and
	\begin{align}
		{{\bf{H}}_{\chi}}\left[ n \right] = {\left. {\frac{{\partial {{\bf{h}}_{\chi}}}}{{\partial {{\bf{s}}_{\chi}}}}} \right|_{{{\bf{s}}_{\chi}} = {{{\bf{\tilde s}}}_{\chi}}\left[ {n\left| {n - 1} \right.} \right]}},
	\end{align}
	respectively. 
	\item The covariance matrix of the posterior MSE is obtained as  
	\begin{align}
		{{\bf{M}}_{\chi}}\left[ {n\left| {n - 1} \right.} \right] &= {{\bf{G}}}\left[ {n - 1} \right]{{\bf{M}}_{\chi}}\left[ {n - 1} \right]{{\bf{G}}^H}\left[ {n - 1} \right] \nonumber \\
		& + {{\bf{Q}}_{\bf{s}}}. \label{predicate_M}
	\end{align}
	\item 	The Kalman gain is obtained as
	\begin{align}
		{{\bf{K}}_{\chi}}\left[ n \right] &= {{\bf{M}}_{\chi}}\left[ {n\left| {n - 1} \right.} \right]{{\bf{H}}_{\chi}}^H\left[ n \right]\nonumber\\
		&{\left( {{{\bf{H}}_{\chi}}\left[ n \right]{{\bf{M}}_{\chi}}\left[ {n\left| {n - 1} \right.} \right]{{\bf{H}}_{\chi}}^H\left[ n \right] + {{\bf{Q}}_m}} \right)^{ - 1}}.\label{Kalman}
	\end{align}
	\item 	The posterior state is obtained as 
	\begin{align}
		{{\bf{\tilde s}}_{\chi}}\left[ n \right] &= {{\bf{K}}_{\chi}}\left[ n \right]\left( {{{\bf{m}}_{\chi}}\left[ n \right] - {{\bf{h}}_{\chi}}\left( {{{{\bf{\tilde s}}}_{\chi}}\left[ {n\left| {n - 1} \right.} \right]} \right)} \right)  \nonumber \\
		& + {{\bf{\tilde s}}_{\chi}}\left[ {n\left| {n - 1} \right.} \right].  \label{posterior_s}
	\end{align}
	\item 	The covariance matrix of ${{\bf{M}}_{\chi}}\left[ n \right]$ is updated as
	\begin{align}
		{{\bf{M}}_{\chi}}\left[ n \right] = \left( {{\bf{I}} - {{\bf{K}}_{\chi}}\left[ n \right]{{\bf{H}}_{\chi}}\left[ n \right]} \right){{\bf{M}}_{\chi}}\left[ {n\left| {n - 1} \right.} \right].  \label{posterior_M}
	\end{align}
	
\end{enumerate}

\subsection{Problem Formulation}
Due to the uncertainty in the target’s location information, the constraint in (\ref{ekf2}), shown at the top of the next page, is adopted in this work to ensure the effectiveness of the beamforming design, 
where $r$ denotes the angular resolution, 
$\theta _n^{{\rm{cov}}}$ and $\phi _n^{{\rm{cov}}}$ denote the angle sets that satisfy the conditions $\left| {{\theta _{B\chi}}[n - 1] - \theta _n^{{\rm{cov}}}} \right| \le L{\sigma _{{\theta _{\chi}}}}\left[ {n - 1} \right]$ and $\left| {{\phi _{B\chi}}[n - 1] - \phi _n^{{\rm{cov}}}} \right| \le L{\sigma _{{\phi _{\chi}}}}\left[ {n - 1} \right]$, respectively. 
According to the basic properties of the Gaussian distribution, we set $L = 3$. Considering the limitation on computational complexity, the angular resolution is set to $r = {0.1^ \circ }$ \cite{ZhouS2024TCom}.
\begin{figure*}[ht]
		\begin{align}
			&\left| {{\bf{a}}_{B{\chi}T}^H\left( {{\theta _{B\chi}}\left[ {n - 1} \right],{\phi _{B\chi}}\left[ {n - 1} \right]} \right){{\bf{W}}_{\chi}}\left[ n \right]} \right. \nonumber\\
			&\quad \quad \quad \quad   \left. { \times {{\bf{a}}_{B{\chi}T}}\left( {{\theta _{B\chi}}\left[ {n - 1} \right],{\phi _{B\chi}}\left[ {n - 1} \right]} \right) - {\bf{a}}_{B{\chi}T}^H\left( {\theta _n^{{\rm{cov}}},\phi _n^{{\rm{cov}}}} \right){{\bf{W}}_{\chi}}\left[ n \right]{{\bf{a}}_{B{\chi}T}}\left( {\theta _n^{{\rm{cov}}},\phi _n^{{\rm{cov}}}} \right)} \right| \le r{\rm{tr}}\left( {{{\bf{W}}_{\chi}}\left[ n \right]} \right), \nonumber\\
			&\quad \quad \quad \quad \quad   \forall \left| {{\theta _{B\chi}}[n - 1] - \theta _n^{{\rm{cov}}}} \right| \le L{\sigma _{{\theta _{\chi}}}}\left[ {n - 1} \right],\forall \left| {{\phi _{B\chi}}[n - 1] - \phi _n^{{\rm{cov}}}} \right| \le L{\sigma _{{\phi _{\chi}}}}\left[ {n - 1} \right]
			\label{ekf2}
		\end{align}
	\hrulefill
\end{figure*}

In this work, the trajectory of $B$ and the beamforming vectors are jointly designed to maximize the achievable SR of $B$, thereby enhancing the secrecy performance of the considered system. 
Let ${\bf{Q}} = \left\{ {{{\bf{q}}_B}\left[ n \right],{{\bf{v}}_B}\left[ n \right]} \right\}$ and ${\bf{W}} = \left\{ {\bf{W}}_U\left[ n \right] = {\bf{w}}_U\left[ n \right]{\bf{w}}_U^H\left[ n \right], {{\bf{W}}_E}\left[ n \right] = {{\bf{w}}_E}\left[ n \right]{\bf{w}}_E^H\left[ n \right] \right\}$, the following optimization problem is formulated
\begin{subequations}\label{Opt}
	\begin{align}
		\mathcal{P}_{0}: &\mathop {\max }\limits_{{\bf{W}},{\bf{Q}}} {R_{\sec }}\left[ n \right]   \\
		{\mathrm{s.t.}}\; 
		& {\rm{tr}}\left( {{{\bf{W}}_U}\left[ n \right] + {{\bf{W}}_E}\left[ n \right]} \right) \le {P_{\max }}, 		\label{P_MAX}\\
		& {{\bf{W}}_U}\left[ n \right]\succcurlyeq0,{{\bf{W}}_E}\left[ n \right]\succcurlyeq0,			\label{tr_W}\\
		&{\rm{rank}}\left( {{{\bf{W}}_U}\left[ n \right]} \right) = 1, {\rm{rank}}\left( {{{\bf{W}}_E}\left[ n \right]} \right) = 1,		\label{RANK}\\
		&{\left| {{\bf{a}}_{B{\chi},T}^H\left( {{\theta _{B{\chi}}}\left[ n \right],{\phi _{B{\chi}}}\left[ n \right]} \right){{\bf{w}}_{\chi}}\left[ n \right]} \right|^2} \ge {\Gamma _{\mathrm{echo}}}, \forall {\chi}, \label{echo_E}\\
		&{{\bf{q}}_B}\left[ {n + 1} \right] - {{\bf{q}}_B}\left[ n \right] = {\delta _t}{{\bf{v}}_B}\left[ n \right],						\label{q_B}\\
		& \left\| {{{\bf{v}}_B}\left[ n \right]} \right\| \le {\rm{V}}_{\max }^{xy},											\label{V_B}\\
		&\left\| {{{\bf{v}}_B}\left[ {n + 1} \right] - {{\bf{v}}_B}\left[ n \right]} \right\| \le a_{\max }^{xy}{\delta _t},	\label{a_B}\\
		&0 \le {P_{{\rm{hor}}}}\left[ n \right] \le P_{\max }^{{\rm{hor}}},								\label{Ph_max}\\
		& \textrm {(\ref{ekf2})}, 	\nonumber
	\end{align}
\end{subequations}
where 
${P_{\max }}$ denotes the maximum transmit power in each time slot, 
${\Gamma _{\mathrm{echo}}}$ signifies the sensing threshold, 
${\rm{V}}_{\max }^{xy}$ and $a_{\max }^{xy}$ denote the maximum tolerable horizontal flight velocity and acceleration thresholds of $B$, respectively, 
and $P_{\max }^{{\rm{hor}}}$ represents the maximum tolerable horizontal flight power of $B$.

Specifically, 
(\ref{P_MAX}) imposes the maximum transmit power limit in each time slot, 
(\ref{tr_W}) and (\ref{RANK}) denote the constraints for the beamforming vectors, 
{(\ref{echo_E}) specifies the sensing requirement for $B$ and $E$,}
(\ref{q_B})–(\ref{a_B}) define the constraints related to the flight velocity of $B$, 
(\ref{Ph_max}) represents the horizontal flight power constraint of $B$, 
and 
(\ref{ekf2}) characterizes the relationship between the sensing error and the beamwidth. 
Essentially, by adjusting the beam gain over the potential angular region, the proposed formulation facilitates the generation of an optimized beam pattern that closely approximates an ideal radar beam pattern with adjustable beamwidth \cite{ZhouS2024TCom,LiJ2007SPM}.

Due to the coupling between the variables ${\bf{Q}}$ and ${\bf{W}}$, as well as the presence of the non-convex constraint in (\ref{Ph_max}) and the fractional form of the objective function, directly solving the original problem ${{\cal P}_0}$ is intractable.

\section{Proposed Solution}
\label{sec:proposed solution}

In this section, alternate optimization approach is adopted, where ${\bf{Q}}$ and ${\bf{W}}$ are optimized alternately in an iterative manner while treating the other variables as fixed. Accordingly, $\mathcal{P}_{0}$ is decomposed into two single-variable subproblems. 
By alternately solving these subproblems, the optimization result in each time slot is driven to converge within a predefined accuracy.

\subsection{Subproblem 1: Beam Optimization}

In this subsection, ${\bf{W}}$ is optimized with given ${\bf{Q}}$. 
Accordingly, ${{\cal P}_0}$ is reformulated as
\begin{subequations}
	\begin{align}
		\mathcal{P}_{1.1}: &\mathop {\max }\limits_{\bf{W}} {R_{\sec }}\left[ n \right] \label{P1.1}\\
		{\mathrm{s.t.}}\; & \textrm {(\ref{ekf2})}, \textrm{(\ref{P_MAX})}-\textrm {(\ref{echo_E})}.  \nonumber
	\end{align}
\end{subequations}
It is worth noting that the objective function in (\ref{P1.1}) is neither convex nor concave, 
therefore, ${{\cal P}_{1.1}}$ cannot be solved directly.
By introducing a slack variable $\eta $, ${{\cal P}_{1.1}}$ is rewritten as
\begin{subequations}
	\begin{align}
		\mathcal{P}_{1.2}: &\mathop {\max }\limits_{{\bf{W}},\eta } \eta   \\
		{\mathrm{s.t.}}\; &{R_{\sec }}\left[ n \right] \ge \eta  \label{R_sec_1}\\
		&\textrm {(\ref{ekf2})}, \textrm{(\ref{P_MAX})}-\textrm {(\ref{echo_E})}.  \nonumber
	\end{align}
\end{subequations}
It is worth noting that constraint (\ref{R_sec_1}) is non-convex with respect to ${\bf{W}}$. 
By introducing the slack variables ${H_1}\left[ n \right]$, ${I_1}\left[ n \right]$, ${H_2}\left[ n \right]$, and ${I_2}\left[ n \right]$, (\ref{R_sec_1}) is rewritten as
\begin{align}     
	{\log _2}\left( {1 + \frac{1}{{{H_1}\left[ n \right]{I_1}\left[ n \right]}}} \right) - {\log _2}\left( {1 + \frac{1}{{{H_2}\left[ n \right]{I_2}\left[ n \right]}}} \right) \ge \eta , \label{R_sec_1.2}
\end{align}
with 
{
\begin{subequations}
	\begin{align}
		{H_1}\left[ n \right] \ge {F_1}\left[ n \right], \label{H1}	\\
		{I_1}\left[ n \right] \ge {F_2}\left[ n \right], \label{I1} 	\\
		{H_2}\left[ n \right] \le {F_3}\left[ n \right], \label{H2}	\\
		{I_2}\left[ n \right] \le {F_4}\left[ n \right], \label{I2}	
	\end{align}
\end{subequations}
where   
${F_1}\left[ n \right] = \frac{{{\rho _0}{\rm{tr}}\left( {{\varpi _U}\left[ n \right]{{\bf{W}}_E}\left[ n \right]} \right)}}{{d_{BU}^2}\left[ n \right]} + \delta _U^2$,
${F_2}\left[ n \right] = \frac{{d_{BU}^2}\left[ n \right]}{{{\rho _0}{\rm{tr}}\left( {{\varpi _U}\left[ n \right]{{\bf{W}}_U}\left[ n \right]} \right)}}$,
${F_3}\left[ n \right] = \frac{{{\rho _0}{\rm{tr}}\left( {{\varpi _E}\left[ n \right]{{\bf{W}}_E}\left[ n \right]} \right)}}{{d_{BE}^2}\left[ n \right]} + \delta _E^2$,
${F_4}\left[ n \right] = \frac{{d_{BE}^2}\left[ n \right]}{{{\rho _0}{\rm{tr}}\left( {{\varpi _E}\left[ n \right]{{\bf{W}}_U}\left[ n \right]} \right)}}$, 
and 
${\varpi _\chi }\left[ n \right]  = {{\bf{a}}_{B\chi,T}}\left( {{\theta _{B\chi}}\left[ n \right],{\phi _{B\chi}}\left[ n \right]} \right){\bf{a}}_{B\chi,T}^H\left( {{\theta _{B\chi}}\left[ n \right],{\phi _{B\chi}}\left[ n \right]} \right)$. 
}

Since the left-hand side (LHS) of (\ref{R_sec_1.2}) is non-convex in ${\bf{W}}$, and the right-hand side (RHS) of the inequality in (\ref{I2}) also involves a function that is convex with respect to ${\bf{W}}$, both (\ref{R_sec_1.2}) and (\ref{I2}) constitute non-convex constraints.

First, based on successive convex approximation (SCA) technology, a lower bound for the LHS of (\ref{R_sec_1.2}) is provided as (\ref{low_R_sec}), shown at the top of the next page, where ${\left( . \right)^{\left( k \right)}}$ denotes a feasible point at the $k$-th iteration.
\begin{figure*}[ht]
	\begin{subequations}
		\begin{align}			
			&{\log _2}\left( {1 + \frac{1}{{{H_1}\left[ n \right]{I_1}\left[ n \right]}}} \right) - {\log _2}\left( {1 + \frac{1}{{{H_2}\left[ n \right]{I_2}\left[ n \right]}}} \right) \ge \nonumber \\
			&{\log _2}\left( {1 + \frac{1}{{H_1^{\left( k \right)}\left[ n \right]I_1^{\left( k \right)}\left[ n \right]}}} \right) - \frac{{{H_1}\left[ n \right] - H_1^{\left( k \right)}\left[ n \right]}}{{\ln 2\left( {H_1^{\left( k \right)}\left[ n \right] + H_1^{\left( k \right)}{{\left[ n \right]}^2}I_1^{\left( k \right)}\left[ n \right]} \right)}} - \frac{{{I_1}\left[ n \right] - I_1^{\left( k \right)}\left[ n \right]}}{{\ln 2\left( {I_1^{\left( k \right)}\left[ n \right] + I_1^{\left( k \right)}{{\left[ n \right]}^2}H_1^{\left( k \right)}\left[ n \right]} \right)}} \nonumber \\
			&- \left\{ {{{\log }_2}\left( {1 + \frac{1}{{H_2^{\left( k \right)}\left[ n \right]I_2^{\left( k \right)}\left[ n \right]}}} \right) - \frac{{{H_2}\left[ n \right] - H_2^{\left( k \right)}\left[ n \right]}}{{\ln 2\left( {H_2^{\left( k \right)}\left[ n \right] + H_2^{\left( k \right)}{{\left[ n \right]}^2}I_2^{\left( k \right)}\left[ n \right]} \right)}} - \frac{{{I_2}\left[ n \right] - I_2^{\left( k \right)}\left[ n \right]}}{{\ln 2\left( {I_2^{\left( k \right)}\left[ n \right] + I_2^{\left( k \right)}{{\left[ n \right]}^2}H_2^{\left( k \right)}\left[ n \right]} \right)}}} \right\}\mathop  = \limits^\Delta  F_5^{\left( k \right)}\left[ n \right]  \label{low_R_sec}			
		\end{align}
		\hrulefill
	\end{subequations}
\end{figure*}
Then, (\ref{I2}) is transformed into (\ref{trans_I2}), which is shown at the top of the next page.
{\begin{figure*}[ht]
	\begin{align}
		{\rho _0}{\mathrm{tr}}\left( {{{\bf{a}}_{BE,T}}\left( {{\theta _{BE}}\left[ n \right],{\phi _{BE}}\left[ n \right]} \right){\bf{a}}_{BE,T}^H\left( {{\theta _{BE}}\left[ n \right],{\phi _{BE}}\left[ n \right]} \right){{\bf{W}}_U}\left[ n \right]} \right) \le \frac{{d_{BE}^2}\left[ n \right]}{{{I_2}\left[ n \right]}}\label{trans_I2}
	\end{align}
	\hrulefill
\end{figure*}}
Thus, ${{\cal P}_{1.2}}$ is reformulated as
\begin{subequations}
	\begin{align}
		\mathcal{P}_{1.3}: &\mathop {\max }\limits_{{\bf{W}},\eta } \eta   \\
		{\mathrm{s.t.}}\; &F_5^{\left( k \right)}\left[ n \right] \ge \eta , \\
		& \textrm {(\ref{ekf2})}, \textrm{(\ref{P_MAX})}- \textrm{(\ref{echo_E})}, \nonumber \\
		& \textrm{(\ref{H1})}- \textrm{(\ref{H2})}, \textrm {(\ref{trans_I2})}. \nonumber
	\end{align}
\end{subequations}

After ignoring the rank-one constraint in (\ref{RANK}), $\mathcal{P}_{1.3}$ becomes a semidefinite programming (SDP) problem, which can be efficiently solved using convex optimization tools such as CVX. Subsequently, a feasible rank-one solution can be recovered by applying techniques such as Gaussian randomization, singular value decomposition (SVD), or other similar methods proposed in \cite{DanQ2025TCCN, DanQ2025TVT}. Moreover, rank-1 matrix reconstruction is performed only once after the overall algorithm converges.

\subsection{Subproblem 2: Trajectory Optimization}
In this subsection, ${\bf{Q}}$ is optimized for given ${\bf{W}}$. Accordingly, ${{\cal P}_0}$ is reformulated as
\begin{subequations}
	\begin{align}
		\mathcal{P}_{2.1}:&\mathop {\max }\limits_{{\bf{Q}},{\bf{\eta }}} \eta   \\
		{\mathrm{s.t.}}\; & \textrm {(\ref{q_B})} - \textrm {(\ref{Ph_max})} ,\textrm {(\ref{R_sec_1})}. \nonumber
	\end{align}
\end{subequations}
It is worth noting that both (\ref{Ph_max}) and (\ref{R_sec_1}) are non-convex with respect to ${\bf{Q}}$; consequently, $\mathcal{P}_{2.1}$ cannot be solved directly.

{First,
${\bf{a}}_{B{\chi},T}^H\left( {{\theta _{B{\chi}}}\left[ n \right],{\phi _{B{\chi}}}\left[ n \right]} \right)$ is iterated by using the trajectories obtained from the previous iteration \cite{LeiH2025IOT,DengC2023TWC}.}
Then by introducing a slack variable ${\lambda _B}\left[ n \right]$, (\ref{Ph_max}) is rewritten as
\begin{subequations}
	\begin{align}
		{P_i}{\lambda _B}\left[ n \right] &+ {P_0}\left( {1 + \frac{{3{{\left\| {{{\bf{v}}_B}\left[ n \right]} \right\|}^2}}}{{U_{{\rm{tip}}}^2}}} \right) \nonumber \\
		& + \frac{1}{2}{d_0}\rho sA{\left\| {{{\bf{v}}_B}\left[ n \right]} \right\|^3} \le P_{\max }^{{\rm{hor}}} \label{P_flight}
	\end{align}
\end{subequations}
with
\begin{align}
	\frac{1}{{{\lambda _B}{{\left[ n \right]}^2}}} \le {\lambda _B}{\left[ n \right]^2} + \frac{{{{\left\| {{{\bf{v}}_B}\left[ n \right]} \right\|}^2}}}{{v_0^2}}.\label{lamada}
\end{align}
Since both sides of the inequality are convex functions, (\ref{lamada}) is non-convex. By replacing the RHS of the inequality with its convex lower bound, (\ref{lamada}) is approximated as  (\ref{lamada_SCA}), shown at the top of this page.
\begin{figure*}[ht]
	\begin{align}
		\frac{1}{{{\lambda _B}{{\left[ n \right]}^2}}} \le \lambda _B^{(k)}{\left[ n \right]^2} + 2\lambda _B^{(k)}\left[ n \right]\left( {{\lambda _B}\left[ n \right] - \lambda _B^{(k)}\left[ n \right]} \right) + \frac{{{{\left\| {{\bf{v}}_B^{(k)}\left[ n \right]} \right\|}^2}}}{{v_0^2}} + \frac{2}{{v_0^2}}{\left( {{\bf{v}}_B^{(k)}\left[ n \right]} \right)^T}\left( {{{\bf{v}}_B}\left[ n \right] - {\bf{v}}_B^{(k)}\left[ n \right]} \right) \label{lamada_SCA} 
	\end{align}
	\hrulefill
\end{figure*}

Next, (\ref{R_sec_1}) is rewritten as 
\begin{align}
	{R_{\sec }}\left[ n \right] &= {\log _2}\left( {1 + \frac{{{C_1}\left[ n \right]}}{{{C_2}\left[ n \right] + d_{BU}^2\left[ n \right]\sigma _U^2}}} \right) \nonumber \\
	& - {\log _2}\left( {1 + \frac{{{C_3}\left[ n \right]}}{{{C_4}\left[ n \right] + d_{BE}^2\left[ n \right]\sigma _E^2}}} \right) \ge \eta, \label{R_sec_SAC1} 
\end{align}
where {
${C_1}\left[ n \right] = {\rho _0}{\rm{tr}}\left( {{\varpi _U}\left[ n \right] {{\bf{W}}_U}\left[ n \right] } \right)$, 
${C_2}\left[ n \right] = {\rho _0}{\rm{tr}}\left( {{\varpi _U}\left[ n \right] {{\bf{W}}_E}\left[ n \right] } \right)$, 
${C_3}\left[ n \right] = {\rho _0}{\rm{tr}}\left( {{\varpi _E}\left[ n \right] {{\bf{W}}_U}\left[ n \right] } \right)$, 
and 
${C_4}\left[ n \right] = {\rho _0}{\rm{tr}}\left( {{\varpi _E}\left[ n \right] {{\bf{W}}_E}\left[ n \right] } \right)$. }
Since the LHS of (\ref{R_sec_SAC1}) is neither convex nor concave, the overall formulation in (\ref{R_sec_SAC1}) becomes non-convex.
By introducing slack variables ${\zeta _1}\left[ n \right]$ and ${\zeta _2}\left[ n \right]$, (\ref{R_sec_SAC1}) is rewritten as
\begin{subequations}
	\begin{align}
		{\log _2}\left( {1 + {C_1}\left[ n \right]{\zeta _1}\left[ n \right]} \right) &- {\log _2}\left( {1 + {C_3}\left[ n \right]{\zeta _2}\left[ n \right]} \right) \ge \eta, \label{R_sec_2}  \\
		{\zeta _1}\left[ n \right] &\le \frac{1}{{{C_2}\left[ n \right] + d_{BU}^2\left[ n \right]\sigma _U^2}}, 	\label{kesai_1}	\\
		{\zeta _2}\left[ n \right] &\ge \frac{1}{{{C_4}\left[ n \right] + d_{BE}^2\left[ n \right]\sigma _E^2}}.	\label{kesai_2}
	\end{align}
\end{subequations}
It should be noted that both (\ref{R_sec_2})-(\ref{kesai_2}) are all non-convex.
Firstly, for (\ref{R_sec_2}), the LHS is replaced by its lower bound (\ref{LHS_R}), which is shown at the top of the next page.
\begin{figure*}[ht]
	\begin{align}
		{\log _2}\left( {1 + {C_1}\left[ n \right]{\zeta _1}\left[ n \right]} \right) &- {\log _2}\left( {1 + {C_3}\left[ n \right]{\zeta _2}\left[ n \right]} \right) \ge \nonumber \\
		&{\log _2}\left( {1 + {C_1}\left[ n \right]{\zeta _1}\left[ n \right]} \right) - \left\{ {{{\log }_2}\left( {1 + {C_3}\left[ n \right]\zeta _2^{\left( k \right)}\left[ n \right]} \right) + \frac{{{C_3}\left[ n \right]\left( {{\zeta _2}\left[ n \right] - \zeta _2^{\left( k \right)}\left[ n \right]} \right)}}{{1 + {C_3}\left[ n \right]\zeta _2^{\left( k \right)}\left[ n \right]}}} \right\}\mathop  = \limits^\Delta  F_6^{\left( k \right)}\left[ n \right]
		\label{LHS_R}
	\end{align}
	\hrulefill
\end{figure*}
Next, (\ref{kesai_1}) and (\ref{kesai_2}) are transformed as
\begin{subequations}
\begin{align}
	{C_2}\left[ n \right] + d_{BU}^2\left[ n \right]\sigma _U^2 \le \frac{1}{{{\zeta _1}\left[ n \right]}}, 	\label{kesai_1_ref}	\\
	{C_4}\left[ n \right] + d_{BE}^2\left[ n \right]\sigma _E^2 \ge \frac{1}{{{\zeta _2}\left[ n \right]}}.	\label{kesai_2_ref}
\end{align}
\end{subequations}
By applying SCA to handle the RHS of (\ref{kesai_1_ref}), (\ref{kesai_1_ref}) is reformulated as
\begin{align}
	{C_2}\left[ n \right] + d_{BU}^2\left[ n \right]\sigma _U^2 \le \frac{1}{{\zeta _1^{\left( k \right)}\left[ n \right]}} - \frac{{\left( {{\zeta _1}\left[ n \right] - \zeta _1^{\left( k \right)}\left[ n \right]} \right)}}{{{{\left( {\zeta _1^{\left( k \right)}\left[ n \right]} \right)}^2}}}. \label{kesai_1_1_ref}
\end{align}
Replacing the LHS with its lower bound, (\ref{kesai_2_ref}) is transformed into (\ref{kesai_2_1_ref}), as shown at the top of the next page.
\begin{figure*}[ht]
	\begin{align}
		{C_4}\left[ n \right] + \sigma _E^2\left[ {{{\left\| {{\bf{q}}_B^{\left( k \right)}\left[ n \right] - {{\bf{q}}_E}\left[ n \right]} \right\|}^2} + 2{{\left( {{\bf{q}}_B^{\left( k \right)}\left[ n \right] - {{\bf{q}}_E}\left[ n \right]} \right)}^T}\left( {{{\bf{q}}_B}\left[ n \right] - {\bf{q}}_B^{\left( k \right)}\left[ n \right]} \right) + z_B^2\left[ n \right]} \right] \ge \frac{1}{{{\zeta _2}\left[ n \right]}} \label{kesai_2_1_ref}
	\end{align}
	\hrulefill
\end{figure*}

Then, 
$\mathcal{P}_{2.1}$ is approximated as
\begin{subequations}
	\begin{align}
		\mathcal{P}_{2.2}:&\mathop {\max }\limits_{{\bf{Q}},\eta ,{\zeta _1},{\zeta _2}} \eta  \\
		{\mathrm{s.t.}}\; 
		&F_6^{\left( k \right)}\left[ n \right] \ge \eta  \label{F6_k}\\
		& \textrm {(\ref{q_B})} - \textrm {(\ref{Ph_max})} ,\textrm {(\ref{P_flight})}, \textrm {(\ref{lamada_SCA})}, \textrm {(\ref{kesai_1_1_ref})},\textrm {(\ref{kesai_2_1_ref})}. \nonumber
	\end{align}
\end{subequations}
$\mathcal{P}_{2.2}$ is a convex optimization formulation that is processed with the CVX toolbox.


{
Finally, an iterative algorithm is proposed based on the block coordinate descent (BCD) method combined with the EKF, which alternately solves the two subproblems $\mathcal{P}_{1.3}$ and $\mathcal{P}_{2.2}$ to approximate a suboptimal solution to the original problem $\mathcal{P}_0$. The optimal solution from the previous time slot is used as the feasible starting point for the current time slot. In each iteration, the current solution serves as the feasible initial point for the next iteration, and the variables are progressively updated via alternating optimization.  
Let $E\left( \mathbf{W}^{(m)}, \mathbf{Q}^{(m)} \right)$ denote the objective value of $\mathcal{P}_0$ at the $m$-th iteration. The detailed iterative procedure is summarized in \textbf{Algorithm 1}, where $\varepsilon_0$ denotes the convergence tolerance.

\subsection{Convergence and Complexity Analysis}

\textbf{Convergence Analysis:}
In step 1 of Algorithm \ref{algorithm1}, a standard linear problem $\mathcal{P}_{1.3}$ is solved and we obtain the solution $\mathbf{W}$. Thus, we have
\begin{align}\label{Al_A}
	E\left( \mathbf{W}^{(m)},\mathbf{Q}^{(m)} \right) \le E\left( \mathbf{W}^{(m+1)},\mathbf{Q}^{(m)} \right).
\end{align}
In Step 2, the solution $\mathbf{Q}^{(m+1)}$ is obtained by solving the convex approximation problem $\mathcal{P}_{2.2}$. We use $E^{\text{lb}}$ to represent the objective function of this approximated problem, which serves as an lower bound of the original objective function in $\mathcal{P}_{0}$. Accordingly, a sequence of inequalities can be derived to verify that the objective value will not rise in this step.
\begin{align}\label{Al_B}
	E\left( \mathbf{W}^{(m)},\mathbf{Q}^{(m)} \right) &\le  E\left( \mathbf{W}^{(m+1)},\mathbf{Q}^{(m)} \right) \notag \\
	&\le E^{\text{ub}} \left( \mathbf{W}^{(m+1)},\mathbf{Q}^{(m+1)}\right) \notag \\
	&\le E\left( \mathbf{W}^{(m+1)},\mathbf{Q}^{(m+1)} \right).
\end{align}

As a result, the objective function of $\mathcal{P}_{0}$ keeps monotonically increasing over the entire iteration procedure. 
Moreover, this function is upper-bounded since the feasible region is determined by the given constraints. 
Therefore, we can conclude that \textbf{Algorithm \ref{algorithm1}} achieves convergence. 

\textbf{Complexity Analysis:}
In each time slot, the EKF involves a matrix inversion operation, whose computational complexity is cubic, i.e., $\mathcal{O}(4^3)$ \cite{WuJ2023TVT}. Let $r_n$ denote the number of iterations for solving (\ref{Opt}) in the $n$-th time slot, satisfying $r_n \leq r_{\max}$, where $r_{\max}$ is a preset maximum iteration count. Among the subproblems of this optimization, $\mathcal{P}_{1.3}$ is an SDP problem with an $M$-antenna UPA, while $\mathcal{P}_{2.2}$ is solved via the SCA method. When a standard interior-point method is employed to solve these convex optimization problems, the computational complexity is given by $\mathcal{O}\left[4^3 + r_n\left( \left( M^2 + 1 \right)^{4.5} + 4^{3.5} \right)\log \left( \frac{1}{\varepsilon_0} \right) \right]$ \cite{LeiH2025IOT}. Therefore, the overall computational complexity of Algorithm \ref{algorithm1} over all time slots is expressed as $\sum_{n = 1}^N \mathcal{O}\left[ 4^3 + r_n\left( \left( M^2 + 1 \right)^{4.5} + 4^{3.5} \right)\log \left( \frac{1}{\varepsilon_0} \right) \right]$.

\begin{algorithm}[tb]
	\caption{Iterative Algorithm for Problem ($\mathcal{P}_{0}$)}\label{algorithm1}
	\KwIn{Set ${\tilde{\mathbf{s}}_{\chi}}[1]$, ${\bf{M}}_{\chi}[1]$, and ${{\bf{q}}_B}[1]$}
	\Do{$n \le N$}
	{
		{
				1. $n = n + 1$;\\
				2. Compute ${\bf{\tilde s}}_{\chi}\left[ {n\left| {n - 1} \right.} \right]$ with (\ref{predicate_s}) and ${\bf{M}}_{\chi}\left[ {n\left| {n - 1} \right.} \right]$ with (\ref{predicate_M});
			}\\
		\Do
		{$E\left( \mathbf{W}^{(m)},\mathbf{Q}^{(m)} \right) - E\left( \mathbf{W}^{(m-1)},\mathbf{Q}^{(m-1)} \right) \succ \varepsilon_0 $}
		{	
				a. Obtain the solution ${\bf W}^{\left( {m + 1} \right)}$ by solving ($\mathcal{P}_{1.3}$) for given $\left\{ 	\mathbf{W}^{(m)},\mathbf{Q}^{(m)} \right\}$;\\
				b. Obtain the solution ${\bf Q}^{\left( {m + 1} \right)}$ by solving ($\mathcal{P}_{2.2}$) for given $\left\{ \mathbf{W}^{(m+1)},\mathbf{Q}^{(m)} \right\}$;\\
				c. $m = m + 1$;\\
				d. Compute the objective value $E\left( \mathbf{W}^{(m)},\mathbf{Q}^{(m)} \right)$.
			}
		{
				4. Compute ${\bf{K}}_{\chi}\left[ n \right]$ with (\ref{Kalman});\\
				5. Utilizing $\mathbf{W}^{(m)}$ and $\mathbf{Q}^{(m)}$ obtain $\mathbf{m}_{\chi}\left[ n \right]$;\\
				6. Utilizing $\mathbf{m}_{\chi}\left[ n \right]$ obtain $\tilde{\mathbf{s}}_{\chi}\left[ n \right]$ and $\mathbf{M}_{\chi}\left[ n \right]$ based on (\ref{posterior_s}) and (\ref{posterior_M});
			}
	}
\end{algorithm}

}

\section{Numerical Results and Analysis}
\label{sec:Simulation}

\begin{table}[tb]
	\caption{{Simulation Parameters}}
	\begin{center}
		\renewcommand{\arraystretch}{1}
		\scalebox{1}{\begin{tabular}{c| c | c| c }
				\Xhline{1.2pt}
				\textbf{Notation}   	& \textbf{Value}  & \textbf{Notation}   	& \textbf{Value}  \\
				\hline			$M$	&  $16$   &   ${f_c}$ &   $30$ GHz \\
				\hline
				${P_{\max }}$ 	&  $30$ dBm   &  $s$  & $0.05$ ${{\textrm m}^3}$ \\
				\hline
				${\delta _t}$		& $0.1$s   & $A$ & $0.503$ ${{\textrm m}^2}$ \\
				\hline
				${\sigma _E^2},\sigma _U^2,\sigma _B^2,$ 	& $-120$ dB, $-120$ dB     	& ${P_{\rm 0}}$ &  $79.86$ W\\
				\hline
				$\sigma _x^2, \sigma _y^2$& $0.5$ m, $0.5$ m  & ${P_{\rm i}}$  & $88.63$ W \\
				\hline
				$\sigma _{vx}^2,\sigma _{vy}^2 $ 	& $0.5$ m/s, $0.5$ m/s &  ${v_0}$&  $4.03$ m/s \\  
				\hline
				${a_i},i = 1,2$	& $6.7e^{-7}$, $2e^{4}$  &  ${U_{\rm tip}}$& $120$ m/s\\
				\hline
				${a_i},i = 3,4$	& $1$, $1$  &  ${d_0}$  & $0.6$ \\
				\hline
				${\rho_0 }$ 	& $-30$ dB  &  $\rho $ & $1.225$ kg/${{\textrm m}^3}$ \\
				\Xhline{1.2pt}
		\end{tabular}}
	\end{center}
	\label{table3}
\end{table}

This section evaluates the performance of the proposed scheme through numerical results. Unless otherwise specified, all simulation parameters are set according to Table \ref{table3} \cite{DengC2023TWC}.

\subsection{Communication and Sensing Performance of the Proposed Scheme}

\begin{figure*}[t]
	\centering
	\subfigure[]{
		\label{fig2a}
		\includegraphics[width = 0.31  \textwidth]{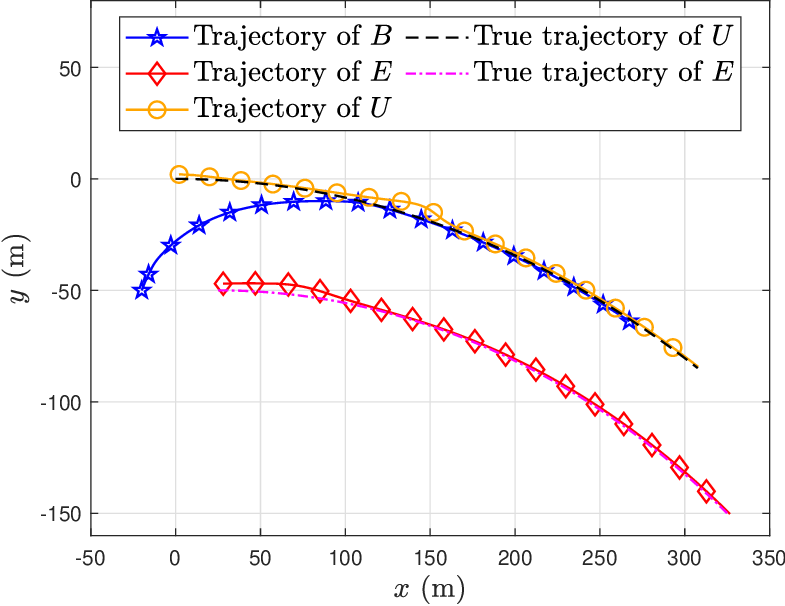}}
	\subfigure[]{
		\label{fig2b}
		\includegraphics[width = 0.31  \textwidth]{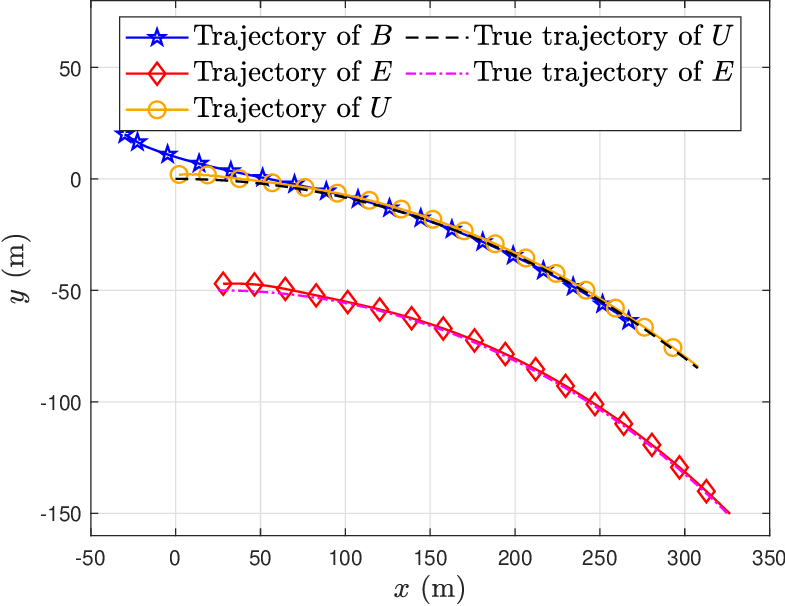}}	
	\subfigure[]{
		\label{fig2c}
		\includegraphics[width = 0.31  \textwidth]{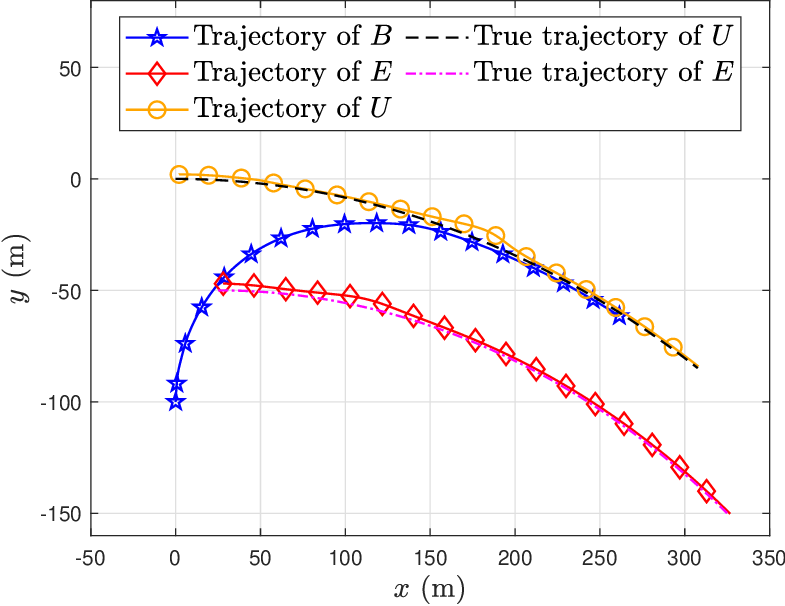}}
	\caption{Trajectories. (a) Starting from $[-20,-50]$. (b) Starting from $[-30, 20]$. (c) Starting from $[0,-100]$.}
	\label{fig02}
\end{figure*}

Fig.~\ref{fig02} illustrates the flight trajectories of $B$ with the proposed scheme for three different initial departure locations, along with the corresponding predicted trajectories of $U$ and $E$. As shown in Figs.~\ref{fig2a}–\ref{fig2c}, $B$ continuously predicts the motion states of both $U$ and $E$. Given that the system objective is to maximize the SR, it can be clearly observed that $B$ gradually flies toward $U$ and persistently tracks its motion to enhance the quality of the legitimate link. Simultaneously, $B$ maintains tracking and jamming toward $E$ to optimize the overall security performance. Moreover, the predicted trajectories of $U$ and $E$ closely align with their true trajectories, exhibiting only minor deviations. This consistency validates the favorable convergence behavior and tracking accuracy of the adopted EKF-based estimation framework.

\begin{figure*}[t]
	\centering
	\subfigure[]{
		\label{fig3a}
		\includegraphics[width = 0.31  \textwidth]{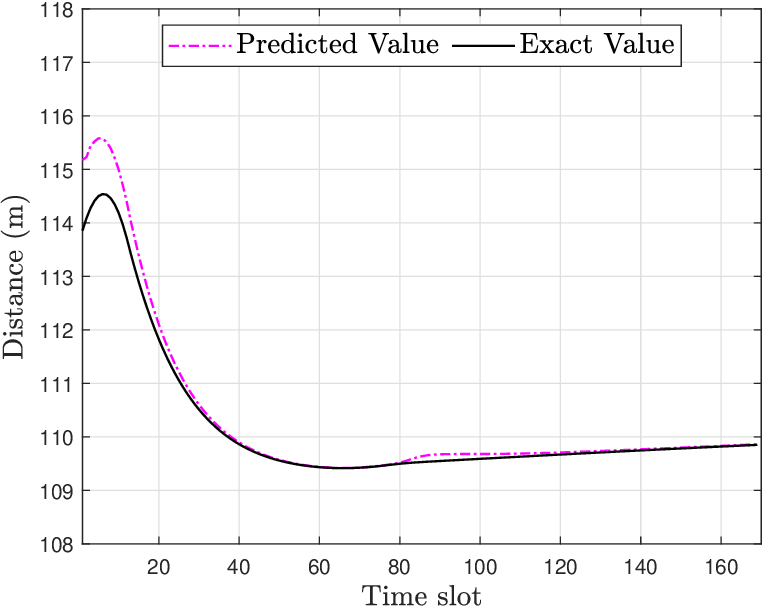}}
	\subfigure[]{
		\label{fig3b}
		\includegraphics[width = 0.31  \textwidth]{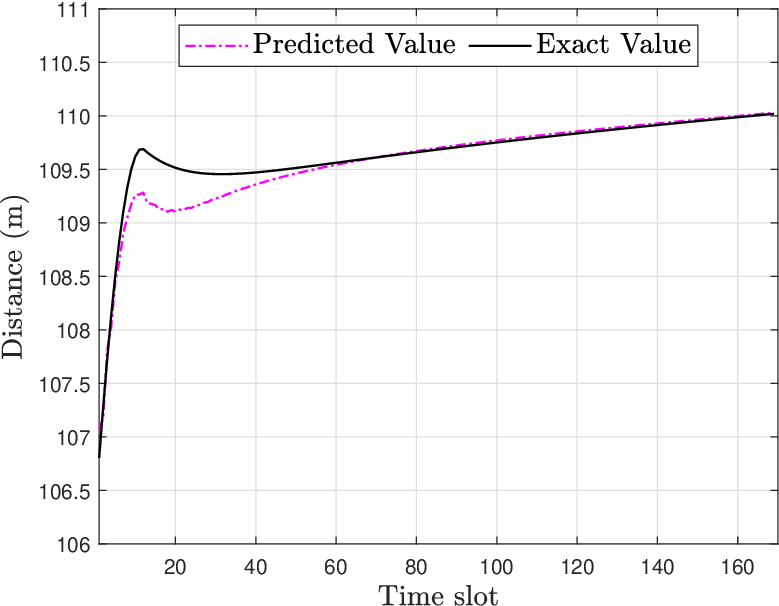}}	
	\subfigure[]{
		\label{fig3c}
		\includegraphics[width = 0.31  \textwidth]{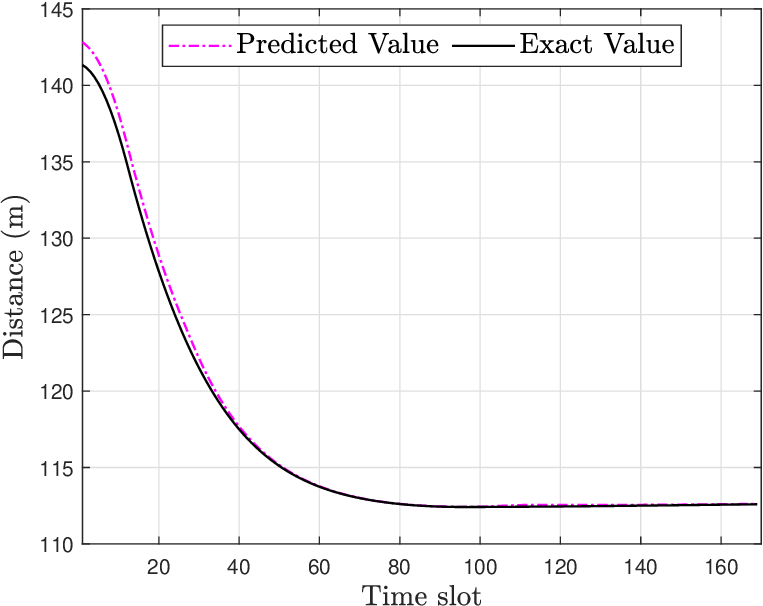}}
	\caption{Distance between $B$ and $U$. (a) Starting from $[-20,-50]$. (b) Starting from $[-30, 20]$. (c) Starting from $[0,-100]$.} 
	\label{fig03}
\end{figure*}

Fig.~\ref{fig03} depicts the distance between the actual and predicted positions of $B$ and $U$ in three different initial location scenarios. As shown in Figs.~\ref{fig3a} and~\ref{fig3c}, the distance between $B$ and $U$ gradually decreases, indicating that $B$ continuously tracks $U$ and progressively reduces the separation.
In Fig.~\ref{fig3b}, however, the distance between $B$ and $U$ exhibits a slight increase during the initial phase. This is primarily attributed to the difference in initial location, combined with the fact that the UAV's initial velocity is zero while its maximum speed matches that of $U$, which temporarily limits its ability to close the gap immediately.
Moreover, across all three scenarios, the error between the predicted distance and the actual distance follows a consistent pattern: it is relatively large during the initial stage and gradually converges over time. This behavior aligns well with the characteristics of the EKF. Specifically, the initial estimation errors stem mainly from uncertainties in the initial state and the linearization approximations inherent in the EKF. As successive measurement updates are incorporated, the filter progressively refines its estimates, approaching the true state and thereby achieving error convergence.

\begin{figure*}[t]
	\centering
	\subfigure[]{
		\label{fig04a}
		\includegraphics[width = 0.31  \textwidth]{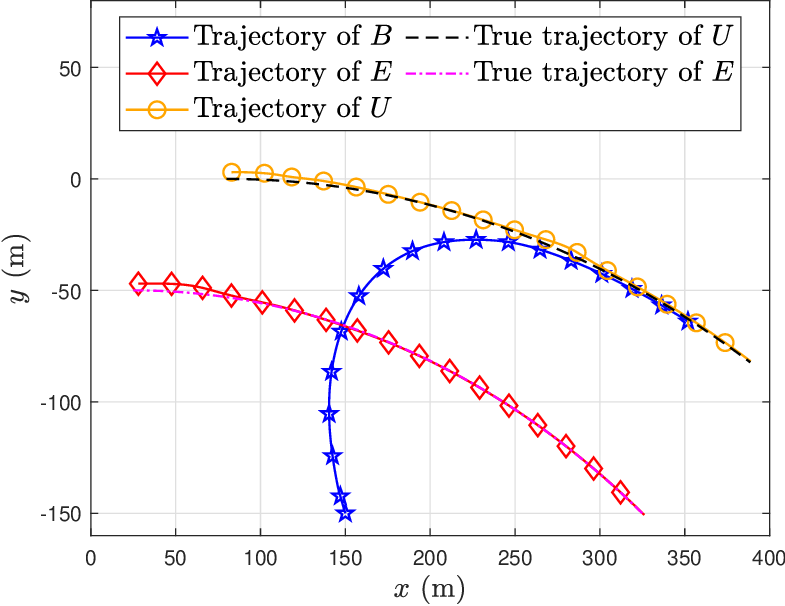}}
	\subfigure[]{
		\label{fig04b}
		\includegraphics[width = 0.31  \textwidth]{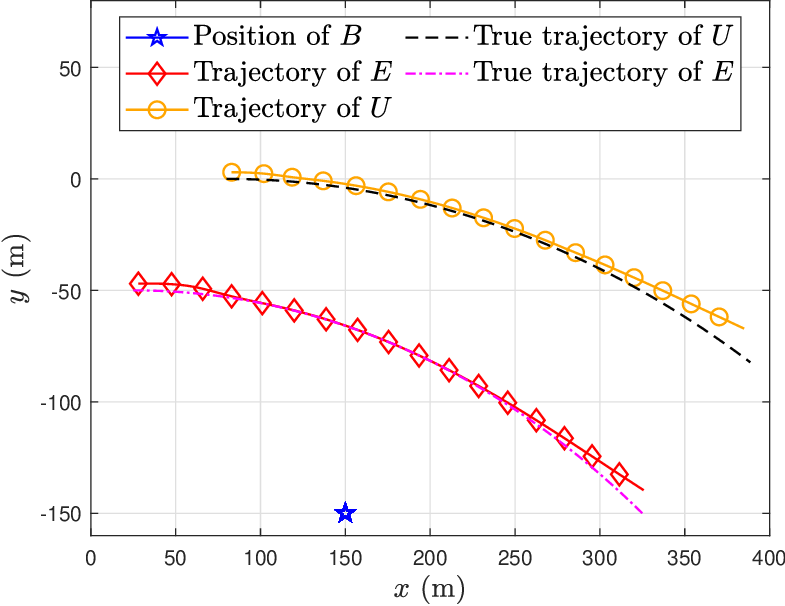}}	
	\subfigure[]{
		\label{fig04c}
		\includegraphics[width = 0.31  \textwidth]{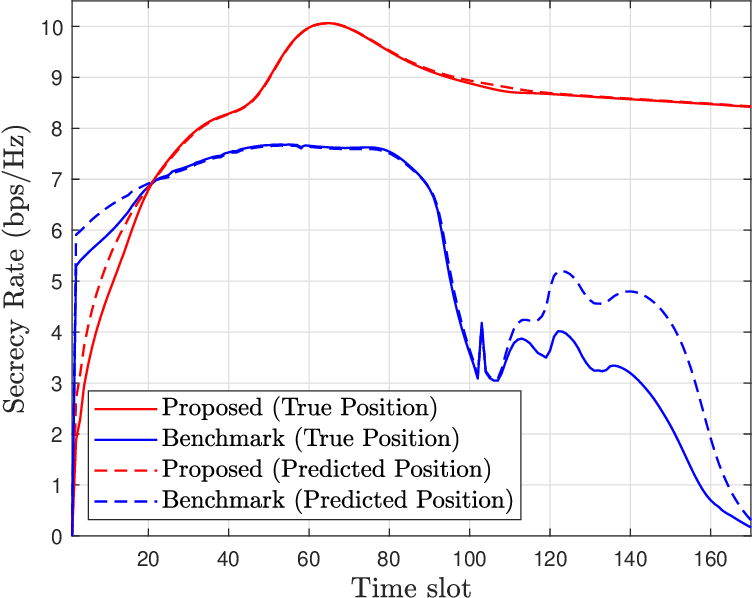}}	
	\caption{Trajectories. (a) Proposed. (b) Benchmark. (c) Secrecy rate.}
	\label{fig04}
\end{figure*}

\begin{figure}[t]
	\centering
	\subfigure[]{
		\label{fig05a}
		\includegraphics[width = 0.31  \textwidth]{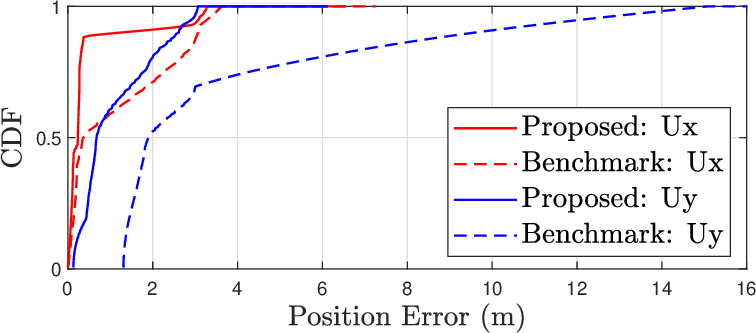}}
	\subfigure[]{
		\label{fig05b}
		\includegraphics[width = 0.31  \textwidth]{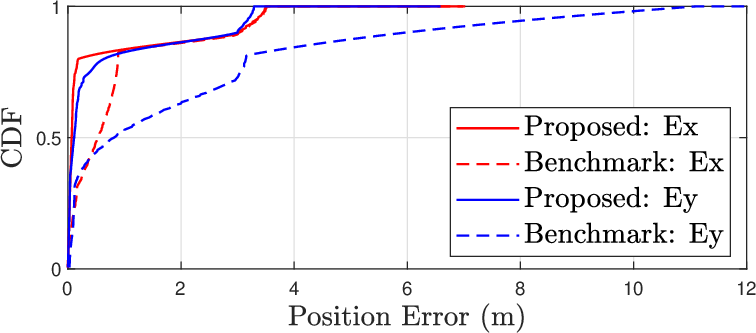}}	
	\caption{CDFs of estimated errors. (a) CDFs of $U$'s position error. (b) CDFs of $E$'s position error.}
	\label{fig05}
\end{figure}

\subsection{Comparison of the Proposed Scheme and Benchmark}

To comprehensively evaluate the system performance, we reconfigure the UAV’s initial position as well as the motion trajectories of $U$ and $E$, and compare the proposed scheme with the following benchmark scheme:
$B$ is deployed at a fixed location $[150, -150]^{T}$ at an altitude of 80 m and is equipped with a multi-antenna array. By optimizing the transmit beamforming, it performs sensing and prediction of the motion states of both $U$ and $E$, following an approach similar to that in \cite{LiuP2023TVT}.

Fig.~\ref{fig04} illustrates the flight trajectory of $B$ under the proposed scheme, starting from the initial position $[150, -150]$, along with the predicted trajectories of $U$ and $E$. It can be observed that in the benchmark scheme, as $U$ and $E$ continue to move, their distance from $B$ gradually increases. This leads to a continuous decrease in the strength of the echo signals received at $B$, which in turn significantly increases the measurement noise. As a result, the EKF-based predictions eventually diverge. In contrast, in the proposed scheme, $B$ is capable of continuously tracking the moving targets, thereby maintaining the stability of the sensing performance and effectively preventing filter divergence.
Fig.~\ref{fig04c} compares the SR achieved by the proposed scheme with that of the benchmark scheme. As expected, the proposed scheme yields a higher SR, benefiting from the synergistic gains obtained through the joint optimization of beamforming and UAV trajectory.
Furthermore, it can be observed that under the proposed scheme, the predicted SR initially exhibits a certain deviation from the actual value but gradually converges as time progresses. In contrast, under the benchmark scheme, the predicted SR tends to diverge from the actual value in the later stage. This trend is consistent with the previously discussed behavior of the EKF-based tracking performance, further validating the stability and robustness of the proposed scheme in dynamic environments.

Fig.~\ref{fig05} compares the cumulative distribution function (CDF) of estimation errors achieved by the proposed scheme and the benchmark scheme. It can be observed that the proposed scheme generally attains higher estimation accuracy than the benchmark scheme.
This improvement can be attributed to the adaptive tracking capability of the UAV in the proposed scheme, which continuously follows the target's motion. In contrast, under the benchmark scheme, the target gradually moves away from the fixed UAV, leading to significant attenuation of the echo signals. Since the strength of the echo signal is closely related to the measurement noise level, a decrease in signal power results in increased measurement noise, thereby introducing larger deviations during the observation process. Under such adverse conditions, the noisy measurements become less effective in correcting the EKF predictions, which may cause the angle estimation errors to accumulate over time.

\subsection{Performance with varying parameters}
\begin{figure*}[t]
	\centering
	\subfigure[]{
		\label{fig06a}
		\includegraphics[width = 0.31  \textwidth]{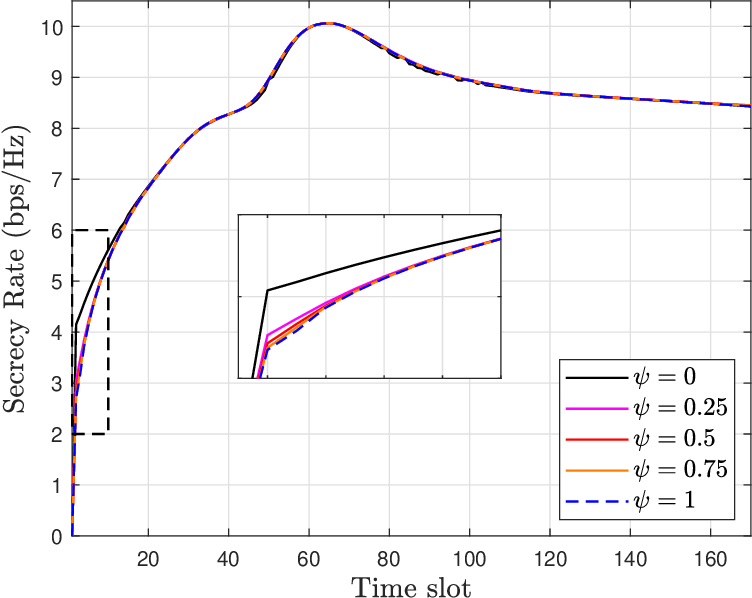}}
	\subfigure[]{
		\label{fig06b}
		\includegraphics[width = 0.31  \textwidth]{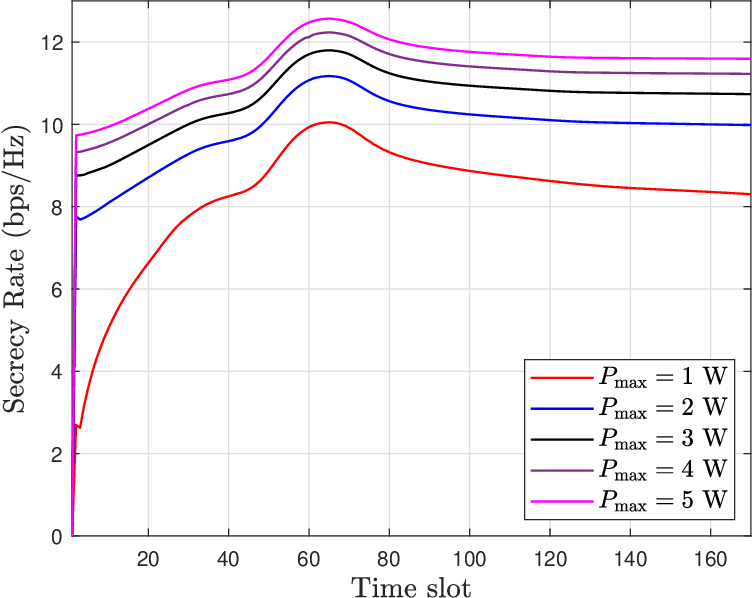}}	
	\subfigure[]{
		\label{fig06c}
		\includegraphics[width = 0.31  \textwidth]{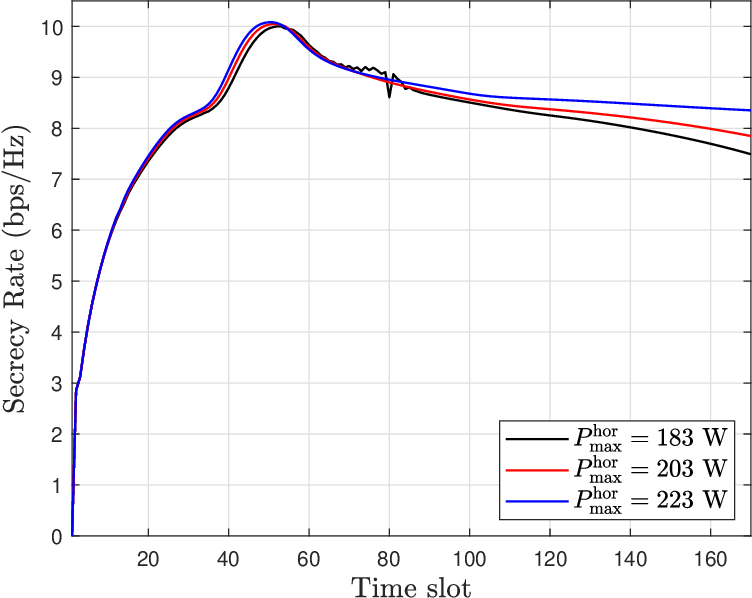}}
	\caption{Secrecy rate for varying parameters. (a) interference residual level $\psi$. (b) the maximum transmit power $P_{\max}$. (c) the maximum flight power  $P_{\max }^{{\rm{hor}}}$.} 
	\label{fig06}
\end{figure*}


{
	Fig.~\ref{fig06a} illustrates the effect of $\psi$ on the achievable secrecy rate. A smaller $\psi$ yields a higher secrecy rate because it imposes weaker interference on the legitimate user, with the secrecy rate maximized at $\psi=0$. Notably, the secrecy rates for $\psi=0.5$ and $\psi=1$ exhibit little difference. This is because when $B$ aligns its beam toward $E$, the signal power directed to $U$ remains inherently low. Consequently, the additional interference caused by varying $\psi$ is marginal, leading to negligible variation in the secrecy rate. 
	Fig.~\ref{fig06b} shows the secrecy rate as a function of $P_{\max}$. It can be seen that the secrecy rate increases with $P_{\max}$, but the growth rate gradually slows down as $P_{\max}$ continues to increase. This is because a higher transmit power improves the SNR at both the legitimate user and the eavesdropper. Consequently, there exists an upper bound on the secrecy rate, as demonstrated in \cite{LeiH2017CL}. 
	Fig.~\ref{fig06c} demonstrates the positive correlation between the secrecy rate and the propulsion power $P_{\max}^{\text{hor}}$. Specifically, a higher $P_{\max}^{\text{hor}}$ enables a faster approach of $B$ towards $U$, which enhances the secrecy rate. Conversely, a lower $P_{\max}^{\text{hor}}$ restricts the maximum flight speed of $B$, increasing the time needed to reach $U$ and consequently degrading the secrecy performance.
}

\section{Conclusion}
\label{sec:Conclusion}

This work investigated a UAV-assisted secure communication system empowered by ISAC technology, where an aerial base station delivers secure communication services to legitimate users while simultaneously jamming a mobile eavesdropper and performing real-time trajectory prediction and tracking for both via an EKF. By jointly optimizing transmit beamforming and UAV trajectory, the achievable secrecy rate was maximized, and an efficient iterative algorithm combining alternate optimization, SCA, and the EKF was developed to address the resulting non-convex problem. {The proposed EKF-based ground-user tracking framework can be extend to aerial-user scenarios via a six-dimensional state vector and a constant-acceleration model to handle higher maneuverability. 
	To further enhance system performance, future work will focus on designing more robust measurement models against clutter and multipath effects and employing multi-UAV coordination to overcome the challenges posed by NLoS conditions. }

\end{document}